\documentclass[12pt,a4paper]{article}
\usepackage{amsmath}
\usepackage{amssymb}
\usepackage{bm}
\usepackage{graphicx}
\usepackage{rotating}
\usepackage{pdfsync}
\usepackage{multirow}
\usepackage{natbib}
\usepackage{har2nat}
\usepackage{pdflscape}
\usepackage{setspace}
\usepackage{xcolor} 
\usepackage[colorlinks,bookmarksopen,bookmarksnumbered,citecolor=red,urlcolor=red]{hyperref}

\title{Calibration of prediction rules for life-time outcomes using prognostic Cox regression survival models and multiple imputations to account for missing predictor data with cross-validatory assessment.}
\author{Mertens, B. J. A.$^1$}

\begin{document}

%
%
%
%

\maketitle
\vspace{-0.5cm} \noindent $^1$Medical Statistics,  Department of Biomedical Data Sciences, Leiden University Medical Center, PO Box 9600, 2300 RC Leiden,  Netherlands
$^2$.
\\
\mbox{}
\\
\noindent {\bf Keywords}  \,\,\, Survival analysis, Multiple imputation, Missing data, Cross-validation, Prediction, Prognosis,  Assessment, Calibration
\date

\section{Introduction}
\subsection{Problem area and state-of-art}
Methodological work on the use of multiple imputation (MI) \cite{Carlin2015} \cite{Carpenter2013} \cite{Rubin1987} to account for missing predictor data  in  both the calibration  and subsequent assessment of prediction rules is underdeveloped.
This applies particularly to prognostic studies and prediction of life-times.
Recent work by Mertens, Banzato and de Wreede (\cite{Mertens2020}) investigates the problem of calibrating prediction rules for binary outcome in the context of two prognostic life-time follow-up studies, when multiple imputations are used to account for the presence of missing observations in predictors.
The authors simplify the prediction problem by reducing the outcome to the observation of death (yes/no) within a fixed window of observation for each patient after study inclusion, as opposed to explicitly modeling the observed survival times.
A general methodological template on the use of multiple imputation in the  calibration of predictive rules is presented which
identifies two basic approaches.
The first consists of averaging the patient-specific predictions which are obtained from separately fitted prediction models on each imputed data set within a sequence of multiple imputed datasets.
We will refer to this approach as prediction-averaging (approach 1).
The second and more traditional approach pools the effect estimates across the distinct regression models fitted on multiple imputations using Rubin's rules, after which the pooled estimate is plugged into the regression equations to define a ``combined model" which is then applied for prediction of the outcome.
We will refer to this method as plug-in pooled regression (approach 2).
The authors implement and evaluate the methodologies for the prediction of early death and demonstrate that there is little difference between
both approaches when predicting binary response in terms of root mean squared error.
Importantly however, substantial gains can be made with respect to the variance induced by imputations and thus the reliability of the final predictions,  when using the first approach.

While \cite{Mertens2020} generate important new insights in the specific application to binary outcome, for full generality and in the context of prognostic studies, explicit modeling of the observed life-times must be addressed. Such generalization of predictive calibration with multiple imputations for missing predictors to survival outcome is however non-trivial and involves at least three fundamental problems.
First, we typically need to account for the presence of censored observation of the outcome when modelling life-times.
Second, Cox models are typically used in clinical application for survival outcome. In the prediction context this implies we must explicitly account for variation in baseline hazards which may be due to imputation variation and can no longer be ignored as nuisance parameters. Thus, we need to expand the parameter space  beyond the usual effect estimates to include the baseline hazard estimates. When using the classical formulation of the Cox model,  the baseline hazards and effect estimates are typically separated within the predictor equation through a non-linear transform and thus no longer on same scales. This  complicates straightforward applications of Rubin's rules.
Finally, multiple imputations themselves need to take the censored nature of the outcomes into account.

\subsection{Objectives and outline of paper}
The objectives of this paper are to extend and adapt  the methods described in \cite{Mertens2020} on the combination of multiple imputation with predictive calibration for binary outcome to the analysis of censored survival outcomes using Cox modeling.
Calibrating a prediction rule on partially observed data and then applying it for the prediction of new observations, which may themselves also contain missing values in the predictors, is a complex statistical problem.
A fundamental problem in addition to the above discussed issues which are specific to life-times, is that while imputation is typically understood to require outcome when estimating missing values (\cite{Moons2006}\cite{Carpenter2013}), these are not available by design on new data for which predictions must be generated.
Similar problems occur with attempts at internal validation,  such as cross-validation for example,   where outcome data corresponding to the items being predicted should  not be used prior to generation of  predictions.   The solution described in this paper allows for joint estimation  of the  prediction models as well as (imputation) of the missing predictor values in new observations.
Conversely,  for internal validation,  we explain how this approach can be combined  with cross-validation, although the methodology is generally applicable.

We   demonstrate and evaluate the proposed methodologies on the same two datasets (CRT and CLL data) previously investigated by \cite{Mertens2020}, but this time removing the simplification to binary outcome for early death such that the full life-time data is analysed, using standard Cox regression models estimated with the partial likelihood.
For ease of reference we commence the paper with a review of the two datasets (section 2).
Section 3 defines methodological extension to survival outcomes.
We describe extensions of the binary outcome methods presented in \cite{Mertens2020} to the censored survival setting, for  both the prediction-averaging based approach (method 1) as well as
plugging-in the pooled estimates (method 2) for Cox models (section \ref{section313}),  which can be implemented using standard imputation software.
Two distinct versions of the plug-in pooling approach for survival analysis are presented.
In addition,  we describe and contrast these methods with na\"ive approaches, where we generate imputations \emph{prior} to  subsequent (cross-validatory) calculations on the full data, using all outcome data, including from validation samples for which predictions are generated.
While the latter approach would clearly be impossible in any ``true" predictive setting where the outcome must still be observed,  the approach is also questionable for internal validation due to potential optimistic bias.
Section 4 applies the respective methods (as well as the na\"ive implementations) in cross-validation on the real data and formulates conclusions with particular reference to root mean squared error and imputation-induced variance of the calibrated predictions.
Cox proportional hazards models are used throughout to calibrate prognostic models, taking censoring into account and regressing on all variables without variable selection, subject to administrative censoring as described further on.
This section also describes the summary performance measures used, with special attention to variation.
Results are presented based on predicted survival probabilities at 1 and 5 years of follow-up.
Section 5 presents a simulation exercise for method comparison and validation.
Various combinations of missing value patterns and strength of association of predictors with outcome are investigated.
We finish with conclusions and a discussion.

\section{Data}
This research arose from our regular clinical collaborations and was specifically motivated through the analysis of two real datasets in clinical survival analysis  studies we encountered in that work.
We therefore present a performance study of the proposed methodologies in a data analytic application in these two datasets. We first introduce the data.

\subsection{CRT data}
The CRT data consists of an observational cardiology cohort of 1053 patients (described in \cite{Hoke}), who underwent CRT (Cardiac Resynchronization Therapy) implantation, with the objective to study patient prognosis using all-cause mortality as outcome (494 deaths out of 1053 patients (47\%), of which 438 are cardiovascular related). The median follow-up is 60 months.  There are 14 predictor variables: \emph{Age}, \emph{Egfr}, \emph{Hb}, \emph{Lvef}, Gender, AtrAct, His, ET, Nyha, Dm, Mr, Qrs, Lbbb, Lvdias. Four of these are continuous (italic script) with the remainder binary or polychotomous. There are 529 patients records containing missing information, of which the majority is concentrated in a single variable (Lvdias) which is missing in 524 cases (50\%). In addition, information is absent for Egfr in 2 cases, for Hb in 7 cases, for Lvef in 20 cases and finally for Mr in 30 cases. In this paper, Cox models are generated on this data adding all 14 predictors as linear effects and using dummy variables in the usual manner, as also done in the original clinical publication \cite{Hoke}. Before analysis, an administrative censoring is applied at 84 months from baseline.

\subsection{CLL data}
The CLL data describe the risk factors and outcomes of a cohort of patients with Chronic Lymphocytic Leukemia (CLL)  who received an allogeneic hematopoietic stem cell transplantation. Data have been extracted from the registry of the European Society for Blood and Marrow Transplantation (EBMT) and have been upgraded and updated through a data quality initiative. The impact on several outcomes of a large series of risk factors, including patient-, disease-, procedure- and center-related information, has been studied in two papers (\cite{Schetelig2017a} and \cite{Schetelig2017b}). In both papers,  multiple imputation was used to account for missing values. The first paper calculates predicted survival and competing risks probabilities for reference patients by pooling coefficients of the Cox models and the baseline hazards. The second paper studies a frailty model for unmeasured center heterogeneity as well as pooled coefficients and  p-values of significance tests of the frailty components.

In the current paper, we use a simplified version of one of the analyses in Schetelig 2017a. The outcome of interest is overall survival (OS) up to five years after first allogeneic stem cell transplantation (data were artificially censored at 5 years). The dataset contains 694 patients for whom 314 deaths were observed. For the whole group, OS was 64\% (SE 2\%) at 2 years and 47\% (SE 2\%) at 5 years. The predictor variables included in the model are age at transplantation (continuous), performance status indicated by the Karnofsky Index (four categories), remission status at transplantation (three categories), cytogenetic abnormalities (four categories), previous autologous transplantation (two categories), donor type (three categories), patient-donor sex match (four categories) and conditioning regimen (three categories). Information is missing for performance status (9\%), remission status (6\%), cytogenetic abnormalities (25\%), patient-donor sex match (1\%) and conditioning regimen (1\%). Administrative censoring was applied at 5 years prior to analysis.

\section{Methodology}

Let $\bm{T}=(T_1,...,T_n)^T$ be a vector of univariate survival time outcomes on  $n$ individuals. With same notation as in \cite{Mertens2020}, we write $\bm{X}=(\bm{X}_m,\bm{X}_o)$ for a corresponding matrix of predictors, consisting of the missing and observed components $\bm{X}_m$ and $\bm{X}_o$ respectively, where we note that missing predictor data may be observed on distinct variables between different patients. In most survival applications,   survival outcome will only be partially known due to censoring,  such that we only observe follow-up times $\bm{T}$ subject to a corresponding vector of binary censoring indicators $\bm{D}$.
The objective is to calibrate a prediction rule based on the calibration data $(\bm{T}, \bm{D}, \bm{X})$ to estimate survival probabilities
$\widehat{S}(t|\widetilde{\bm{X}})$
at times $t$ for future observations with predictors
$\widetilde{\bm{X}}$ for which outcome is not yet observed.
As for the calibration data,   samples to which the predictor rule is applied may themselves contain missing predictor values, such that $\widetilde{\bm{X}}=(\widetilde{\bm{X}}_m,\widetilde{\bm{X}}_o)$.
In the below method description, we initially ignore the issue of the missing data $\widetilde{\bm{X}}_m$ to describe two distinct approaches to predictive calibration in section 3.1.1. We then return to the problem of missing predictor values for new observations in section \ref{section311}.

\subsection{Calibrating Cox regression-based survival predictions with multiple imputations}
We  consider survival predictions calculated from the Cox model
\[
h(t)=h_0(t)\exp{({\bm X\beta})}
\]
with both $h_0(t)$ and ${\bm \beta}$ unknown parameters of the model. We write
${\bm \gamma}=\{{\bm h_0}, {\bm \beta}\}$ for the joint set of model parameters.
When the objective is to apply the model for prediction on new observations,  there are at least two basic approaches to the calibration of survival probabilities when multiple imputations are used to account for missing predictor values.
The first one consists of averaging predicted probabilities from distinct models calculated on single imputations within a sequence of multiple imputed datasets, while the second is application of a Cox model obtained from (Rubin's rules) pooled estimators across multiple imputations, which are plugged into the model equations.
We explain these in turn, starting with the last approach.

\subsubsection{Two basic approaches to calibration with imputation}\label{section311}
In both the predictive averaging and plug-in pooling approaches, we first apply multiple imputation to generate a set of $K$  imputed datasets and calculate separate Cox models on each to obtain the corresponding set of within-model estimates  ${\widehat{\bm\gamma}}_k$, $k=1,...,K$.

The plug-in pooling approach first summarizes the set of separate model estimates to a combined estimate  ${\widehat{\bm\gamma}}_{MI}$ which is plugged into the Cox regression equation to obtain the model $\widehat{h}(t)=\widehat{h}_{0_{MI}}(t)\exp{({\bm X\widehat{\bm \beta}_{MI}})}$, with $\widehat{h}_{0_{MI}}(t)$ and $\widehat{\bm \beta}_{MI}$ `suitably pooled' baseline and hazard ratio estimates based on multiple imputation. The required predictions are then generated from the usual corresponding equations
\[
\widehat{S}(t|\widetilde{\bm{X}})=\widehat{S}_{0_{MI}}(t)^{\exp{(\widetilde{\bm{X}}\widehat{\bm \beta}_{MI})}},
\]
where $\widehat{H}_{0_{MI}}=-\ln(\widehat{S}_{0_{MI}}(t))$ is the estimated cumulative hazard at time $t$.

In contrast, the prediction-averaging method first calculates the within-model predictions
\[
\widehat{S}_k(t|\widetilde{\bm{X}})=\widehat{S}_{0k}(t)^{\exp{(\widetilde{\bm{X}}\widehat{\bm \beta}}_k)},
\]
corresponding to the individual calibrated Cox models $\widehat{h}_k(t)=\widehat{h}_{0k}(t)\exp{({\bm X\widehat{\bm \beta}}_k)}$ on each imputation within the sequence of multiple imputations, for $k=1,...,K$. We then directly aggregate the set of predictions $\{\widehat{S}_k(t|\widetilde{\bm{X}}), k=1,...,K\}$ to a combined estimate $\widehat{S}(t|\widetilde{\bm{X}})$, which may be obtained using means, medians or other suitable statistic.

From a Bayesian methodological  standpoint,  the plug-in pooled approach can be understood as approximating
sampling from the conditional densities
\[
p({\bm \gamma}|\bm{T},\bm{D},\bm{X}_0)=E_{p(\bm{X}_m|\bm{T},\bm{D},\bm{X}_o)}\{ p({\bm \gamma}|\bm{T},\bm{D},\bm{X}_m,\bm{X}_o)  \},
\]
which reveals MI as stochastic integration by first sampling realisations $\widehat{\bm{X}}_{m,k}$ from $p(\bm{X}_m|\bm{T},\bm{D},\bm{X}_o)$ and then $\widehat{\bm \gamma}_k$ from
$p({\bm \gamma}|\bm{T},\bm{D},\widehat{\bm{X}}_{m,k},\bm{X}_o)$ (see \cite{Carpenter2013}, section 2.4 e.g.). Averaging the samples $\widehat{\bm \gamma}_k$ gives rise to the so-called Rubin's rules.
We note that for Cox regression,  a simplistic and direct application of Rubin's rules may be ill-advised due to the presence of the baseline hazards ${\bm h}_0$ in ${\bm \gamma}$.

The prediction-averaging approach in contrast, circumvents calculation of pooled parameter estimates by directly estimating the future outcome $\widetilde{T}$ from the predictive density  by adding an additional level of integration to calculate
\[
\begin{split}
p(\widetilde{T}|\widetilde{\bm{X}},\bm{T},\bm{D},\bm{X}_0)=\int p(\widetilde{T}|\widetilde{\bm{X}},&{\bm \gamma}, \bm{T},\bm{D},\bm{X}_m,\bm{X}_o) p({\bm \gamma}|\bm{T},\bm{D},\bm{X}_m \bm{X}_o)\\
&p(\bm{X}_m |\bm{T},\bm{D},\bm{X}_o)d{\bm \gamma}d\bm{X}_m.
\end{split}
\]
As for `classical' applications of MI for the estimation of model (effect) parameters, stochastic integration may again be used
to approximate the relevant moments of this density
by first sampling the $\widehat{\bm{X}}_{m,k}$ and $\widehat{\bm \gamma}_k$ as described previously  and
then generating either samples $\widetilde{T}_k$ or the expectations $\widehat{S}_k(t|\widetilde{\bm{X}},\widehat{\bm{X}}_{m,k}, \widehat{\bm \gamma}_k)$ corresponding to the conditional densities $p(\widetilde{T}|\widetilde{\bm{X}},\widehat{\bm{X}}_{m,k}, \widehat{\bm \gamma}_k)$ for $k=1,...,K$. In this paper we will pursue the latter option.

\subsubsection{Accounting for missing values in prediction}
For a new observation of which the outcome has not yet been observed and to which the prediction rule will be applied, the corresponding predictor vector $\widetilde{\bm{X}}$
will itself usually contain missing values $\widetilde{\bm{X}}_m$. These must also be estimated.
If multiple imputations were generated in a fully Bayesian approach, such as with Gibbs sampling e.g.,
this could in principle be achieved by generating the required values through simulation from the sequence of conditional densities which describe the joint density.
With such a solution, the required samplers could be determined using the calibration data only and then applied to any new predictor record containing missing data, using suitable starting values for the missing values.
The above paradigm is however associated with substantive practical problems in (survival) application.
\begin{enumerate}
  \item First,  multiple imputation software (such as `\textit{MICE}'  e.g. \cite{vanBuuren2015}) typically is not of the above form and furthermore does not allow for such application to future observations, as
  implementations usually only allow generating the so-called Rubin's-rule pooled estimates.
  \item  A fully  Bayesian  MCMC-based modeling is generally  more difficult to combine with model validation in a predictive sense on set-aside validation data and this computational problem becomes particularly acute for within-sample cross-validation.
  \item For survival analysis particularly, classical implementation of Cox regression using the partial likelihood is cumbersome to combine with the fully Bayesian approach, as described by \cite{Spiegelhalter1996}, page 52, \cite{Clayton1994} or \cite{Ibrahim2001}, chapter 3,  among others.
\end{enumerate}
In analogy to \cite{Mertens2020}, a simpler approach to combine classical partial-likelihood-based Cox modeling with imputation in a predictive scenario where outcomes are not yet available for new observations and with standard software, is to combine the existing calibration data with the new observed predictor data
$\widetilde{\bm{X}}_o$ for which predictions are to be generated.
We then generate multiple imputations $(\widehat{\bm{X}}_{m,k},\widehat{\widetilde{\bm{X}}}_{m,k})$, $k=1,...,K$ for both the unobserved calibration and validation data on this combined dataset
$(\bm{T},\bm{D},\bm{X}_o,\widetilde{\bm{X}}_o)$,  treating the as yet unknown and to be predicted future $\widetilde{T}$ as missing data.  The resulting imputations  $\widehat{\bm{X}}_{m,k}$, $k=1,...,K$ together with the observed calibration data $(\bm{T},\bm{D},\bm{X}_o)$, are then used to fit the corresponding Cox models $\widehat{h}_k(t)=\widehat{h}_{0k}(t)\exp{({\bm X\  \widehat{\bm \beta}}_k)}$ across all imputations. From a technical point of view, this is easily implemented by augmenting the matrix of calibration data with additional rows containing the new predictor data, but with the corresponding outcomes as  missing values. Standard imputation software,  such as `\textit{MICE}' ,  can then be applied on the augmented data matrix,  after which models may be calibrated on the calibration portion of the imputed data. These  models are then applied to the imputed validation portions for prediction.

As discussed above,  we may now either pool the model parameters in some sense across the K models (using Rubin's rules e.g.) to define a combined consensus Cox model and apply this to the K imputed predictor sets $\widehat{\widetilde{\bm{X}}}_k=(\widehat{\widetilde{\bm{X}}}_{m,k},\widetilde{\bm{X}}_o)$, $k=1,...,K$.
Alternatively,  each $k^{th}$ Cox model may be directly applied to its compatible imputed predictor set $\widehat{\widetilde{\bm{X}}}_k$ and the corresponding prediction
$\widehat{S}_k(t|\widehat{\widetilde{\bm{X}}}_k)$ calculated.

Note that,  \emph{irrespective of the approach taken} and when we have missing values in the predictor set,  \emph{there will in full generality be $K$
predictions} $\widehat{S}_k(t|\widehat{\widetilde{\bm{X}}}_k)$, $k=1,...,K$,  although these will coincide for the plug-in pooled approach when predicting fully observed records.
This set of predictions may either be reported as such or summarized using statistics.

\subsubsection{Implementation issues for Cox regression}\label{section313}
To implement the proposed methodologies, we must address two complicating issues for survival analysis generally and  Cox regression specifically.  The first is the \emph{precise} definition of pooling of model parameters with Cox regression  or when aggregating predicted probabilities, in the above description. The second concerns the manner of accounting for censored survival outcomes with imputation.

A special issue arises for the second approach (plug-in pooled method) when we combine the model parameters to generate the pooled model. We have at least two basic approaches available,  denoted 2A and 2B respectively. The first and simplest solution (2A) consists  of a straightforward averaging (Rubin's rules) of the regression parameters and baseline hazard estimates separately,  to define the consensus final model using
\[
  \widehat{h}_{0_{MI}}(t)= \sum_{k=1}^{K} \widehat{h}_{0k}(t)/K
  \quad \textrm{and} \quad
  \widehat{\bm \beta}_{MI}=\sum_{k=1}^{K} \widehat{\bm \beta}_k/K
\]
The second (2B) applies Rubin's rule for the combination of the regression parameters only,  and then use $\widehat{\bm \beta}_{MI}$ to re-estimate the baseline hazard (with the Breslow estimator e.g.). Since we must also take into account the imputation of the missing calibration data, there are again several options to do so.  We decided to take the simplest implementation and define the Breslow estimator using the means of
$\sum_{i=1}^K\widehat{\bm{X}}_{m,k}/K$ of the imputed calibration data.

There are also several options to combine the constituent predictions $\widehat{S}_k(t|\widehat{\widetilde{\bm{X}}}_k)$, $k=1,...,K$ (whether using the first or the second approach),  using Rubin's rules averaging,  applying medians or transform the probabilities using a linearizing transform such as the
logit (e.g.) and then average on the transformed scale.  In this paper we used means-averaging directly on the calculated probabilities.

To account for the presence of censoring, we applied the approach described by \cite{Carpenter2013} (section 8) by substituting the Nelson-Aalen estimates of the cumulative hazard for the observed follow-up time in the calibration data. Imputation models are then constructed using the Nelson-Aalen estimate and censoring indicator,  in addition to the regression covariates. For imputations,  we used the package `\textit{MICE}'  \cite{vanBuuren2011}, which implements the so-called ``chained equations approach", throughout.
With the above choices, we have defined 3 implementations for Cox proportional hazards modeling,  which we refer to in tables, graphs and the remainder of the paper as approaches 1 (prediction-averaging), 2A and 2B (plug-in pooled estimation).

\subsection{Validating a prognostic rule with multiple imputations}
\begin{figure}[!htbh]
\centering
\includegraphics[height=20cm]{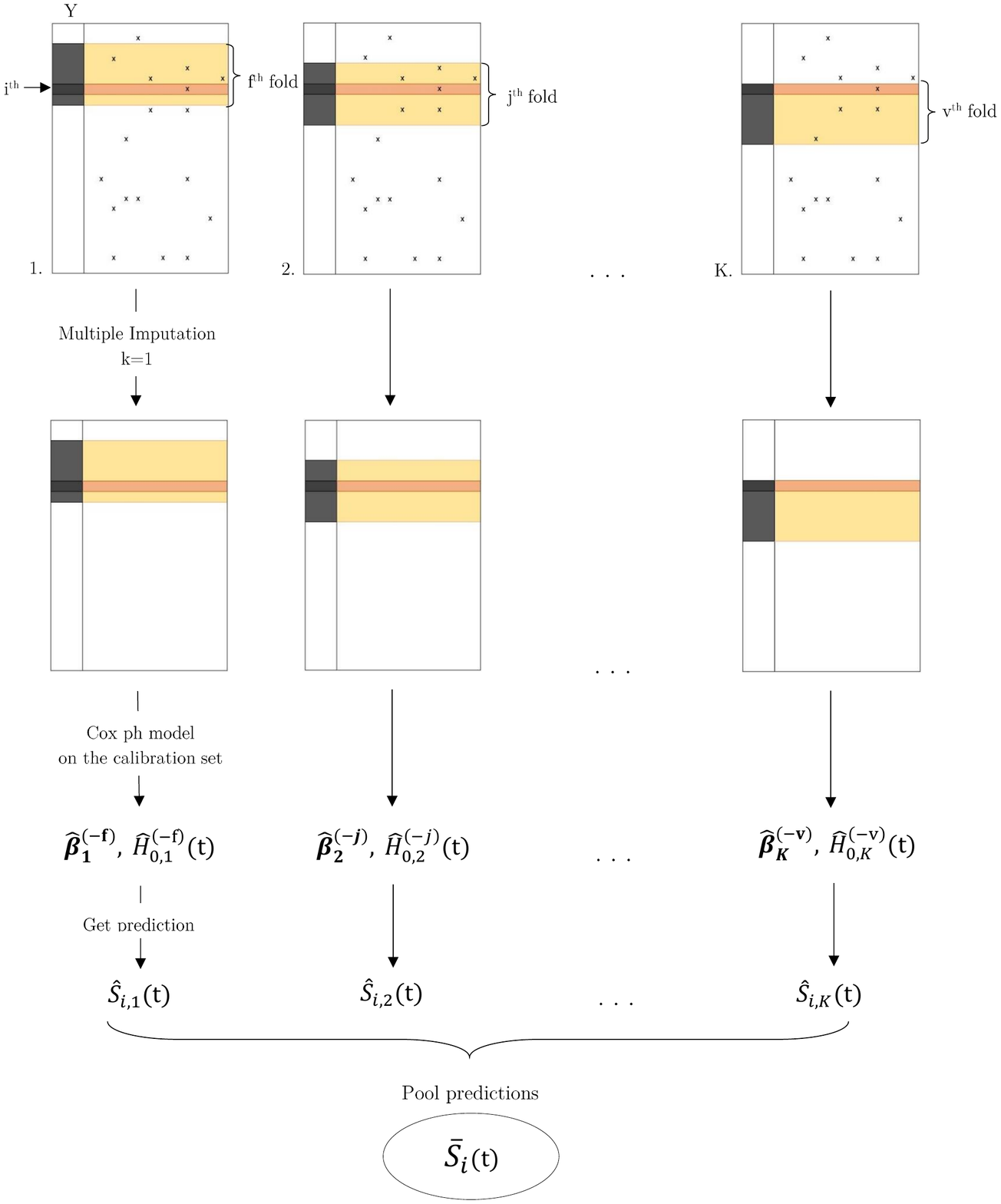}
\caption{\label{fig1} Schematic diagram representing cross-validatory calculations for the prediction-averaging method (approach 1). A fold-partition is defined on the data, after which each fold is taken in turn and a single imputation generated on all data with outcomes removed in the left-out fold. The Cox model is generated on the complementary calibration data and applied to the imputed left-out fold data.  This process is repeated $K$ times for $K$ imputations.  The set of $K$ predictions are combined within individual.
}
\end{figure}

\begin{figure}[!htbh]
\centering
\includegraphics[height=20cm]{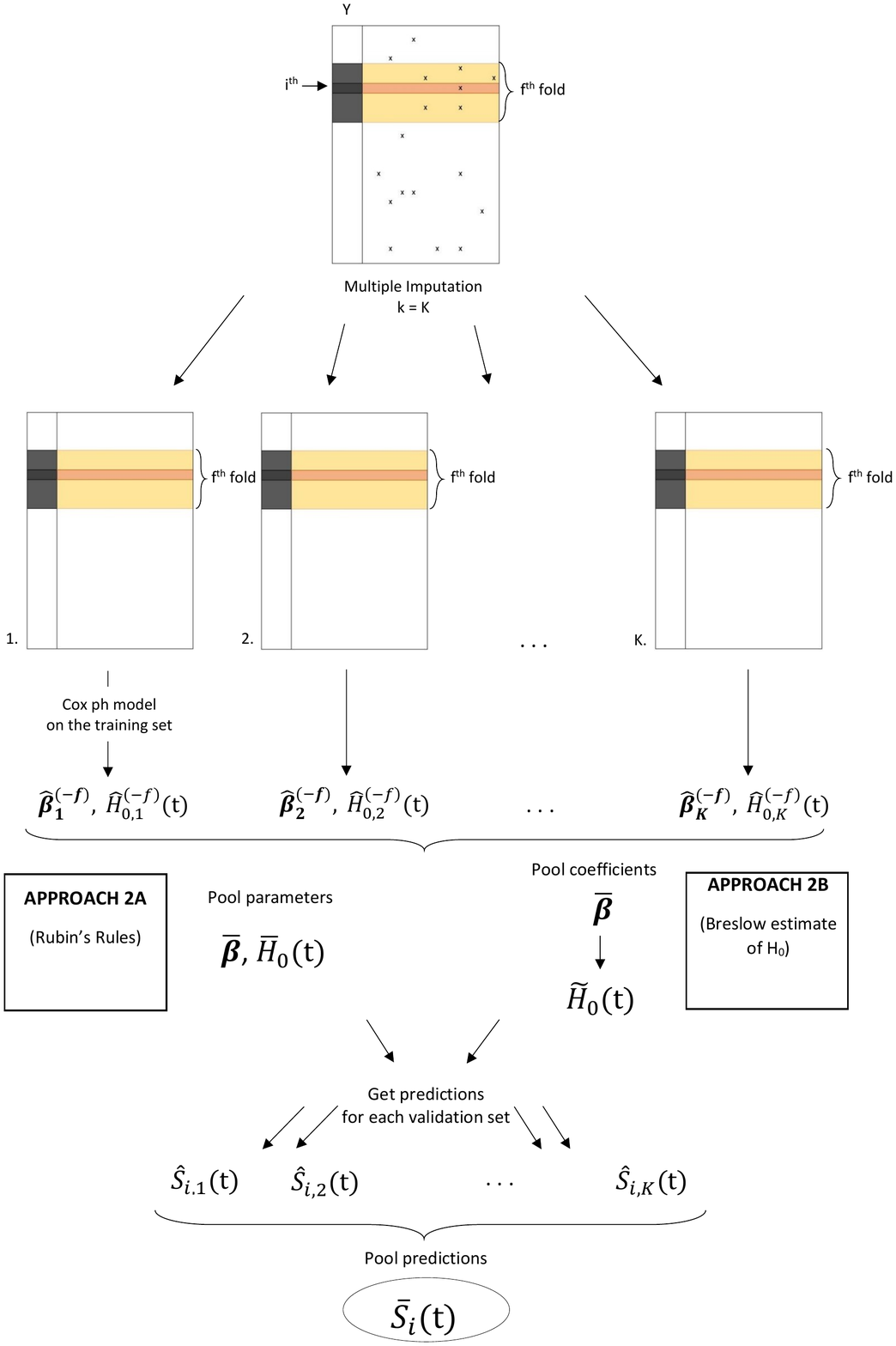}
\caption{\label{fig2}Schematic diagram representing  cross-validatory calculations for the plug-in pooled method (approaches 2A and 2B).
A fold-partition is defined on the data and then kept fixed for the remainder of calculations. $K$ imputations are then directly calculated
on the full data with outcomes removed from each left-out fold. Cox models are then fitted in the calibration part of the data and Rubin's rules are used to generate a pooled model which is applied to the left-out fold imputations.}
\end{figure}

An additional advantage of the above methodology is that it is relatively straightforward to apply with internal validation, such as cross-validation for example.
Once a fold-partition consisting of $L$ folds has been defined on a dataset, we may remove each fold in turn and set its outcome data to missing,  after which the above approaches are applied using the remainder of the data for calibration and apply the prediction rules (1, 2A or 2B) on the (outcome-delected) left-out fold.
The only key difference between approaches 1 and 2 is that the fold definition can be redefined in the first method from each imputation to the next.
Figures~\ref{fig1} and \ref{fig2} highlight the structure of the methods graphically.
Note how for approach 1 we define the fold partition first,  after which imputations are directly generated from the data with outcomes set to missing in the left-out fold. Once the Cox model is fitted on the calibration portion of the data, the predictions may be calculated directly for the left-out data.
Once predictions have been obtained for all observations across all folds in this manner, a new fold definition may be used and the entire process repeated up to a maximum $K$ times.
For approach 2 however,  the fold definition must be kept fixed and $K$ imputations are directly generated on  data with outcomes set missing for each left-out fold.
The pooled model is then estimated across the imputations and only then applied to the left-out fold.

It is worth pointing out that in such cross-validatory calculations,  the fold definitions are completely at random  and thus  the outcome data which is set to missing in any left-out fold is completely at random also, by definition of the procedure.  As explained by Little \cite{Little1992} page 1227, missing outcome data does not contribute to the likelihood equations of the regression of outcome on predictors.

Regarding the Nelson-Aalen estimator calculations and to respect the cross-validatory logic,  these are always computed from the data in the calibration set corresponding to a left-out fold only. The Cox models themselves are subsequently fitted using the original outcome data within the imputed calibration sets.

In the remainder of the paper we restrict to cross-validation,  though as explained the ideas from section 3.1 are quite generally applicable. We implemented the approaches shown in figures~\ref{fig1} and \ref{fig2} using multiple imputations generated with the chained equations methodology implemented in the statistical analysis software R \cite{R2019} using the package `\textit{MICE}'  with standard settings(see \cite{vanBuuren2011} and \cite{vanBuuren2015}, page 276). Cox regression models were fit with the Coxph function from the ``Survival" package \cite{Therneau2019}. The methods presented in this paper have also been implemented in a survival extension of the \textit{R} package  `\textit{mipred}', available
from CRAN and Github (\url{https://github.com/BartJAMertens}).

\subsection{Na\"ive approaches }
We also define so-called ``na\"ive"  implementations of the above 3 approaches.  These consist of
directly computing a set of $K$ multiple imputations on the complete data (including validation outcomes). Once all instances of the $K$ imputed datasets have been obtained, we  define $L$ folds for each imputed dataset separately,  after which the above described process of calculating predictions for the set-aside data in the left-out folds is repeated in analogy to approach 1.
This procedure incorrectly  disregards any subsequent cross-validation or predictive model calibration.
 The process is repeated separately and with  a different fold definition for each of the imputed datasets,  after which we again average predictions within each individual across the imputations. This provides the analogue to the first approach.   The alternative is to again use a fixed fold definition across all $K$ imputed datasets and repeat the procedures for approach 2,  after which Rubin's rule is applied as described to derive the consensus model.

The obvious drawback of the na\"ive procedures is that imputations have full access to the outcome which is to be predicted subsequently and prior to calibration of the models. Likewise,  the Nelson-Aalen estimator used to generate these imputations is calculated on the complete data matrix, in contrast to the previously described methodology.
We refer to the supplementary materials for descriptions of all procedures in pseudo-code format.

\section{Application and assessment in the CRT and CLL data}

In this section, we investigate accuracy (Brier score) and  variability of cross-validated predictions generated with approaches 1 and 2 (A and B) for the CRT and CLL data.
First, we discuss variation between the individual predictions $\widehat{S}_k(t|\widehat{\widetilde{\bm{X}}}_k)$, for $K$ set to 1000 and $t=5$.
We then study the variation between different calibrations of the combined predictions $\overline{S}(t|\widetilde{\bm{X}})$,  when the process is repeated for a new set of imputations and for distinct choices of $K$ ranging from 1 to 10, 100 and finally 1000 imputations and at both $t=1$ and $t=5$ years.
We finally investigate accuracy measures.
All statistics and summary measures are calculated on the output from the $L$-fold cross-validatory approaches described above and in figures~\ref{fig1} and \ref{fig2},  with $L=10$.

\subsection{Variation of individual constituent predictions within the imputation-based approaches}

Figures~\ref{fig3} and \ref{fig4} plot cross-validated survival predictions $\widehat{S}_{i,k}(t)$ at $t=5$ years versus the averages $\overline{S(t)}_i$  for all 3 methods and $K=1000$ imputations, where we introduce $i=1,..,n$ to denote the specific sample with $n$ sample size.  We  distinguish between predictions corresponding to patients with partially missing predictor records (red) and those on fully observed records (black).

\subsubsection{Single imputation-based prediction  (Approaches 1, 2A, 2B)}\label{section411}
We first discuss the uttermost left-side graphs in both figures which plots the $\widehat{S}_{i,k}(t)$ versus the average of these predictions (as defined in approach 1, prediction averaging). Clearly, the variation in individual predictions is high. Observations containing missing values have  higher variation of associated predictions as compared to complete records.

It is important to realize that when $K=1$,  approaches 1, 2A and 2B all coincide and thus, we may interpret
the spread in predictions shown in these left-hand side graphs as prediction variation due to imputation {\it when only a \underline{single imputation} would be used to calibrate a Cox model}.
Following \cite{Mertens2020}, we may calculate a simple summary measure of this variation by calculating deviations  $D(t)_{ik}=\widehat{S}_{i,k}(t)-\overline{S(t)}_i$ around the mean,  which will be heteroscedastic,  but nevertheless of near-constant variance for patients with average survival rate   $\overline{S(t)}_i$ between 0.2 and 0.8.
An ad hoc statistic may now be defined directly at the probability scale, by discarding all deviations corresponding to patients with $\overline{S(t)}_i<0.2$ or $\overline{S(t)}_i>0.8$ and computing the $90^{th}$ and $10^{th}$ percentiles $Q_{0.9}$ and $Q_{0.10}$ across the remaining deviations $D(t)_{ik}$. Finally, we report $R(t)=(Q_{0.9}-Q_{0.10})\times 100$ as a measure of spread of predictive probabilities induced by imputation variation (expressed as percentage).
For the CRT data and when $K=1$, this measure is as large as 15.3\% at 5 years (11.8\% at 1 year) when calculated using predictions from partially observed records only.
These numbers are much smaller when predicting for individuals with fully observed records, respectively 6.7\% and 6.1\% at 5 and 1 years.
We find similar results for the CLL data with $K=1$ as $R=17.3$\% and 13.8\% at 5 and 1 years respectively for partially observed records,   reducing to 9.9\% and 7.5\% for fully observed records respectively. These numbers may also be found in the third column from the right in tables~\ref{table5} and \ref{table6} [(variation of) ``Individual predictions"] and is duplicated in the left-most column of these tables [``Averaged predictions across multiple imputations" (for $K=1$)].
The lower levels of variation for prediction from fully observed records can be explained by imputations inflating variation of predicted probabilities in two ways.  This happens first, due to the induced uncertainty in the fitted models. Furthermore,  when predicting records which contain missing values,  the imputed values will themselves add to the variation of the predicted probabilities.
Crucially, these results suggest single-imputation in the predictive calibration of prognostic rules should  be treated with caution and might be  ill-advised. We return to this point later, when comparing with the multiple-imputation based approaches.

\subsubsection{Prediction using a single pooled model (Approaches 2A, 2B)}
\begin{figure}[!htbh]
\centering
\includegraphics[height=7cm]{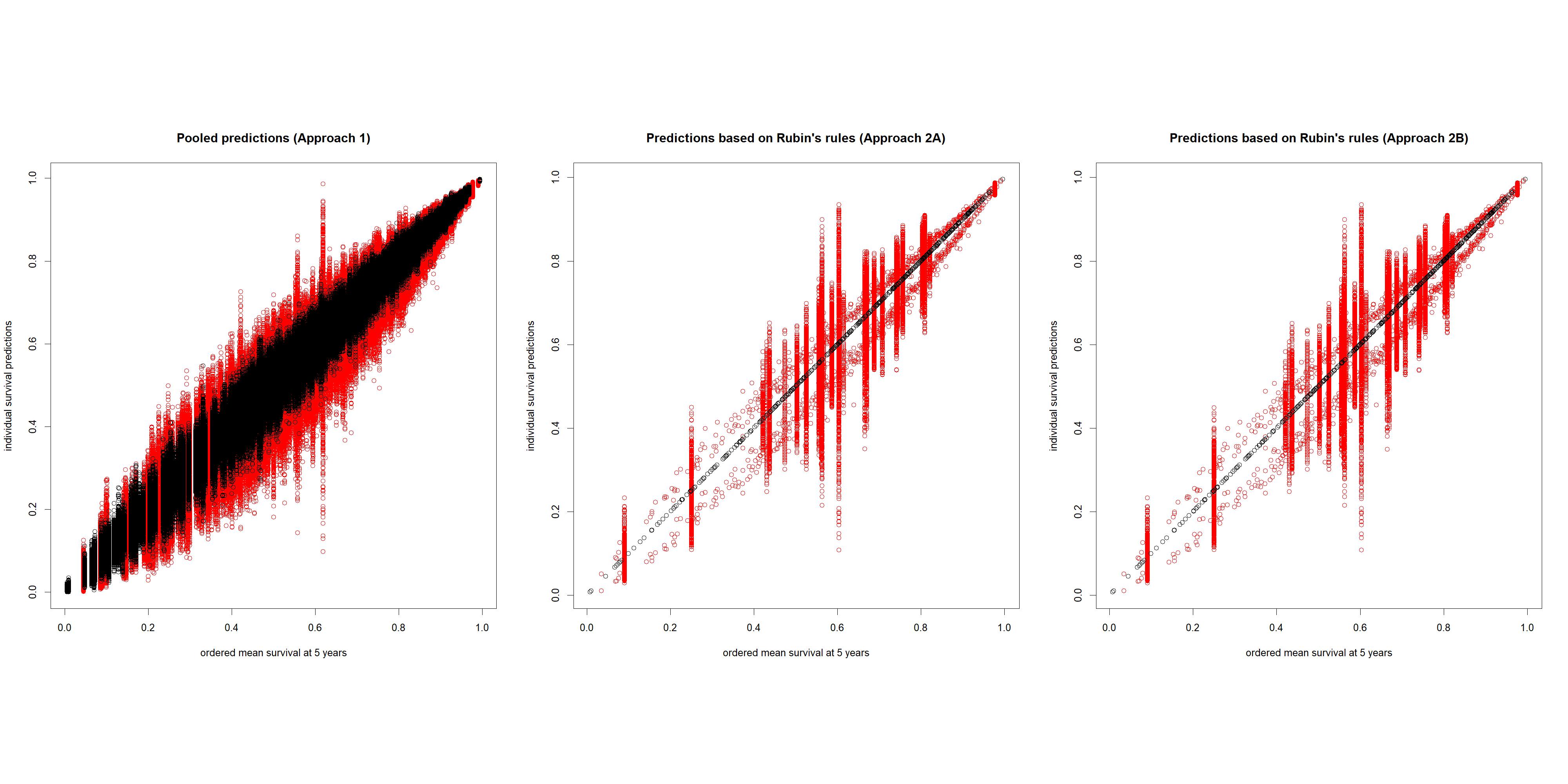}
\caption{\label{fig3}Cross-validated survival predictions $\widehat{S}_{i,k}(t)$ at five years follow-up for the CRT data  within the combined prediction approaches on  multiple imputation   versus the mean predictions $\overline{S(t)}_i$. Results are shown for a single application of approaches 1, 2A and 2B, based on $K=$1000 multiple imputations.  Red plotting symbols show predictions for individuals with missing covariates, while black denotes predictions based on fully observed records.}
\end{figure}

\begin{figure}[!htbh]
\centering
\includegraphics[height=7cm]{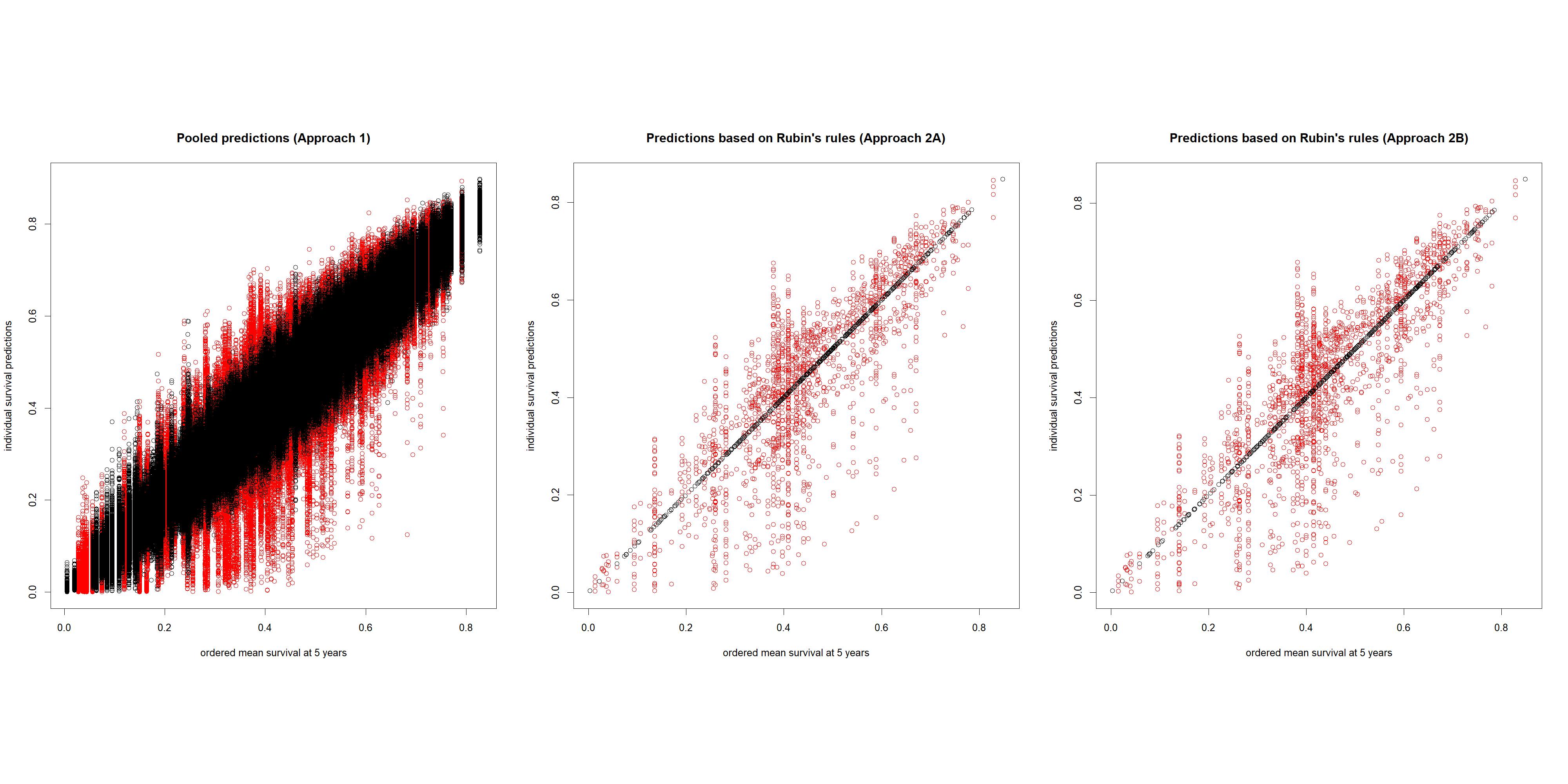}
\caption{\label{fig4}Cross-validated survival predictions $\widehat{S}_{i,k}(t)$ at five years follow-up for the CLL data  within the combined prediction approaches on  multiple imputation  versus the mean predictions $\overline{S(t)}_i$. Results are shown for a single application of approaches 1, 2A and 2B, based on $K=$1000 multiple imputations.  Red plotting symbols show predictions for individuals with missing covariates, while black denotes predictions based on fully observed records.}
\end{figure}

The two plots to the right in figures~\ref{fig3} and \ref{fig4} correspond to approaches 2A and 2B and {\it cannot} be understood as replicates of predictions when $K=1$. Rather they show the individual predictions $\widehat{S}_{i,k}(t)$ generated when a single pooled Cox model is calibrated from Rubin's rules and based on  $K=1000$ imputations. As we are now displaying predictions from a single model, in contrast to subsection \ref{section411}, there is no variation within a patient when predicting from a fully observed record with either implementations 2A or 2B, by design of the approach (predictions follow the diagonal).  Predictions will not coincide for partially observed records,  as these require imputations to be generated and thus we will typically have 1000 predictions for these observations also,  even when using a single pooled Cox model for prediction.  Calculating the `R' variation measure on these predictions as discussed above for approaches 2A and 2B gives results shown in the 2 most right-side columns of tables~\ref{table5} and \ref{table6} [``Individual predictions"]. Note that although a pooled model is used to generate predictions, variation levels are still very large and equal 13.4\% and 10.1\% at 5 and 1 years follow-up for the CRT data. Likewise, these numbers are 14.0\% and 11.2\% for the CLL data.  It can also be seen,  both from the figures and these summary measures that there hardly seems any difference in the variation measures generated for approaches 2A and 2B. This is due to the predictions being virtually identical between both approaches. Results change negligibly when using medians to pool predictions.


Finally, variation of individual predictions is systematically larger at 5 years as compared to 1 year, as would be expected with increasing follow-up.
The compatible results to the above presented in tables \ref{table5} and \ref{table6} for  na\"ive methods show smaller variation measures as compared to the full implementations (especially on the CLL data), but not very pronounced.
This effect could be due to bias in predictive assessment, caused by complete separation between imputation and cross-validation steps within these implementations.

\begin{sidewaystable}[htb]
\normalsize
\begin{center}
\begin{tabular}{ll|r|rrr|rrr|rrr|r|rrr}
\hline
\hline
year&observation&\multicolumn{10}{|l|}{Averaged predictions across multiple imputations}&&\multicolumn{3}{r}{Individual predictions}\\
\cline{3-16}
&status&\multicolumn{1}{|l|}{K=1}&\multicolumn{3}{|l|}{K=10}&\multicolumn{3}{l}{K=100}&\multicolumn{3}{|l|}{K=1000}&&\multicolumn{3}{|r}{K=1000}\\
\cline{1-16}\\
&\multicolumn{1}{l}{}&\multicolumn{12}{l}{Combination of MI and CV:  Approach 1, 2A or 2B}\\
\cline{3-16}
&&ap1,2AB&ap1&ap2A&ap2B&ap1&ap2A&ap2B&ap1&ap2A&ap2B&&ap1&ap2A&ap2B\\ \hline
1 & missing & 11.8 & 3.5 & 5.0 & 5.1 & 1.2 & 4.2 & 4.3 & 0.4 & 3.8 & 3.8 &  & 11.8 & 10.1 & 10.1 \\
  & observed & 6.1 & 1.9 & 5.2 & 5.2 & 0.6 & 4.9 & 4.9 & 0.2 & 5.0 & 5.0 &  & 6.1 & 0.0 & 0.0 \\
5 & missing & 15.3 & 4.6 & 6.5 & 6.4 & 1.5 & 5.0 & 5.0 & 0.5 & 4.7 & 4.7 &  & 15.3 & 13.4 & 13.4 \\
  & observed & 6.7 & 2.1 & 5.2 & 5.2 & 0.7 & 4.9 & 4.9 & 0.2 & 4.9 & 4.9 &  & 6.7 & 0.0 & 0.0 \\
\hline\\
&\multicolumn{1}{l}{}&\multicolumn{12}{l}{Na\"ive combination of MI and CV:  Approach 1, 2A or 2B}\\
\cline{3-16}
&&nv1,2AB&nv1&nv2A&nv2B&nv1&nv2A&nv2B&nv1&nv2A&nv2B&&nv1&nv2A&nv2B\\ \hline
1 & missing & 11.4 & 3.5 & 5.3 & 5.3 & 1.2 & 4.1 & 4.1 & 0.4 & 4.4 & 4.4 &  & 11.4 & 10.1 & 10.1 \\
  & observed & 6.0 & 1.9 & 5.0 & 5.1 & 0.6 & 4.9 & 4.9 & 0.2 & 4.6 & 4.6 &  & 6.0 & 0.0 & 0.0 \\
5 & missing & 15.0 & 4.7 & 6.4 & 6.4 & 1.5 & 5.0 & 5.0 & 0.5 & 4.8 & 4.8 &  & 15.0 & 13.1 & 13.1 \\
  & observed & 6.5 & 1.9 & 5.0 & 5.0 & 0.6 & 4.9 & 4.9 & 0.2 & 4.9 & 4.9 &  & 6.5 & 0.0 & 0.0 \\
\hline
\hline
\end{tabular}
\caption{\normalsize\label{table5} Variance measures $R(t)$ for $t=$ 1 or 5 years calculated on cross-validated prediction for the CRT data. Please refer to the paper for the precise definition of the measure.  The columns ``Individual predictions" show variation between predictions within a single calibration of any approach using $K=1000$.  The columns ``Averaged predictions" show variation between replicate analyses using the same method for either $K=1, 10, 100$ or $1000$. The shorthand notations ``ap1", ``ap2A" and ``ap2B" refer to approaches 1, 2A and 2B respectively, while ``nv1", ``nv2A" and ``nv2B" denote the complementary na\"ive implantations of these.
 }\end{center}
\end{sidewaystable}

\begin{sidewaystable}[htb]
\normalsize
\begin{center}
\begin{tabular}{ll|r|rrr|rrr|rrr|r|rrr}
\hline
\hline
year&observation&\multicolumn{10}{|l|}{Averaged predictions across multiple imputations}&&\multicolumn{3}{r}{Individual predictions}\\
\cline{3-16}
&status&\multicolumn{1}{|l|}{K=1}&\multicolumn{3}{|l|}{K=10}&\multicolumn{3}{l}{K=100}&\multicolumn{3}{|l|}{K=1000}&&\multicolumn{3}{|r}{K=1000}\\
\cline{1-16}\\
&\multicolumn{1}{l}{}&\multicolumn{12}{l}{Combination of MI and CV:  Approach 1, 2A or 2B}\\
\cline{3-16}
&&ap1,2AB&ap1&ap2A&ap2B&ap1&ap2A&ap2B&ap1&ap2A&ap2B&&ap1&ap2A&ap2B\\ \hline
1 & missing & 13.8 & 4.5 & 6.5 & 6.5 & 1.4 & 5.0 & 5.0 & 0.4 & 5.2 & 5.2 &  & 13.8 & 11.2 & 11.2 \\
  & observed & 7.5 & 2.3 & 5.8 & 5.8 & 0.7 & 5.5 & 5.5 & 0.2 & 5.5 & 5.5 &  & 7.5 & 0.0 & 0.0 \\
5 & missing & 17.3 & 5.5 & 8.3 & 8.3 & 1.7 & 6.4 & 6.4 & 0.5 & 6.4 & 6.5 &  & 17.3 & 14.0 & 14.0 \\
  & observed & 9.9 & 3.0 & 7.5 & 7.6 & 1.0 & 7.2 & 7.2 & 0.3 & 7.3 & 7.3 &  & 9.9 & 0.0 & 0.0 \\
\hline\\
&\multicolumn{1}{l}{}&\multicolumn{12}{l}{Na\"ive combination of MI and CV}\\
\cline{3-16}
&&nv1,2AB&nv1&nv2A&nv2B&nv1&nv2A&nv2B&nv1&nv2A&nv2B&&nv1&nv2A&nv2B\\ \hline
1 & missing & 13.5 & 4.3 & 6.3 & 6.3 & 1.4 & 5.2 & 5.2 & 0.4 & 4.9 & 4.9 &  & 13.5 & 11.2 & 11.2 \\
  & observed & 7.1 & 2.2 & 5.3 & 5.3 & 0.7 & 5.3 & 5.3 & 0.2 & 5.1 & 5.1 &  & 7.1 & 0.0 & 0.0 \\
5 & missing & 16.9 & 5.3 & 7.8 & 7.8 & 1.7 & 6.4 & 6.4 & 0.5 & 6.1 & 6.1 &  & 16.9 & 14.1 & 14.1 \\
  & observed & 9.4 & 3.0 & 6.9 & 6.9 & 0.9 & 6.9 & 6.9 & 0.3 & 6.7 & 6.7 &  & 9.4 & 0.0 & 0.0 \\
\hline
\hline
\end{tabular}
\caption{\normalsize\label{table6} Variance measures $R(t)$ for $t=$ 1 or 5 years in prediction for the CLL data. Please refer to the paper for the precise definition of the measure.  The column "individual prediction" shows variation between predictions within a single calibration of any approach using $K=1000$.  The column "averaged predictions" shows variation between replicate analyses using the same method for either $K=1, 10, 100$ or $1000$.
The shorthand notations ``ap1", ``ap2A" and ``ap2B" refer to approaches 1, 2A and 2B respectively, while ``nv1", ``nv2A" and ``nv2B" denote the complementary na\"ive implantations of these.
 }\end{center}
\end{sidewaystable}

\subsection{Exploring variation between replicates of multiple-imputation-based  predictions}

Results from the previous subsection pose a number of questions. One is whether the high levels of predictive variation associated with use of  single imputation  may be reduced by increasing the number of imputations.  Similarly, the absence of variation between the constituent predictions
$\widehat{S}_{i,k}(t)$ in a single calibration of approaches 2A and 2B for fully observed records does however not imply that the final combined predictions $\overline{S(t)}_i$ would remain unchanged -  even for fully observed records - if we were to calibrate the predictive rule anew using a completely different set of imputations and for the same value of $K$,  when $K>1$.

We therefor proceed with an analysis of the variation of the combined predictions $\overline{S(t)}_i$, whether obtained from approach 1 or methods 2A or 2B,  when predictions are compared between  \underline{\it repeated calibrations} of the methods (replications of the analysis) using a new set of imputations for each calibration.
We generated 20 replications of predictive calibrations with the above 3 methodologies, for $K=$ 10, 100 and 1000, in addition to the above discussed results for $K=1$ (single imputation) where all methods coincide. This allows us to investigate the sensitivity of prediction at the patient-level to imputation variation when $K>1$. We discuss results for variation first and continue with an analysis of accuracy in the following subsection.

Let $\widehat{S}(t)_{ir}$ be the predicted (combined) survival probability with any of the above methods at time $t$ for the $i^{th}$ patient and $r$ denotes the replicate (calibration) of the model (to simplify the notation,  we omit the chosen number of imputations used in the calibration).
In full analogy to the above procedure and \cite{Mertens2020}, we now compute variation measures $R(t)$ based on the deviations $D(t)_{ir}=\widehat{S}(t)_{ir}-\overline{S(t)}_i$ at $t=1$ and $t=5$ years and $\overline{S(t)}_i$ the mean across replicates.

Tables~\ref{table5} and \ref{table6} show the calculated statistics $R(t)$  under column ``Averaged predictions across multiple imputations"  for $K=10, 100$ and $1000$ and all 3 methods separately, in addition to the results for $K=1$ (already discussed). These numbers represent ranges of the prediction variation expected at the individual (patient) level should we recalibrate any approach,  due to  variation in multiple imputation.  Multiple conclusions can be drawn.
First, approach 1 has smaller between-replicates variation in prediction as compared to methods 2A or 2B for $K>1$.
This is true,  irrespective of the number of multiple imputations considered (10 to 1000),  both for the prediction from fully observed and with (partially) missing records and at 1 and 5 years follow-up.
Secondly, we notice an at least doubling of the measures $R(t)$ when comparing approaches 2A and 2B with method 1,  for $K=10$ and on fully observed observations,  while numbers increase by a third on partially observed records.
The differences are even larger at $K=100$ and particularly for 1000 imputations where the range statistics `R' are separated by an order of magnitude in favor of approach 1 and irrespective of whether we are predicting on fully or partially observed records.

Thirdly, we find that  variation reduces as we increase the number of imputations on which the prediction combination is based and for all methods.
Tables indicate that 10 imputations are insufficient for predictive calibration in the presence of missing data in these examples and that substantial improvements can be made by increasing this number to 100 at least.
However,  for approach 1,  reductions in predictive variation are still achieved when using 1000 instead of 100 imputations.
Indeed, up to 1000 imputations may be required,  at least for the data investigated in this paper,  as only for $K=1000$ and approach 1 does the predictive variation drop to below 1\%, which seems required for practical use in  medical application.
In contrast, variation does not reduce substantially for  approaches 2A and 2B when increasing imputation numbers beyond 100.
The predictive variation measure $R(t)$ is stuck well above 4.5\% for methods 2A and 2B and for both $K=1000$ and $100$.
Remarkably,  this is also true for prediction based on the fully observed records.
There is little difference in between-replicates variation for approaches 2A and 2B between prediction based on fully observed or partially missing records when $K=1000$.
This  suggests there may be a bound on the number of imputations beyond which the Rubin's rules based methods  no longer profit from increased numbers of imputations.
Predictive variation is substantively and systematically lower for fully observed records only for approach 1,  when comparing with results for partially missing records.
The latter feature would be expected, as predicting from partially missing records needs to account for the additional uncertainty due to the required imputation.
The above results are consistent across  analyses of both the CRT and CLL data,  suggesting results represent a general property of the methods and not the individual data.

\begin{figure}[!htbh]
\centering
\includegraphics[height=7cm]{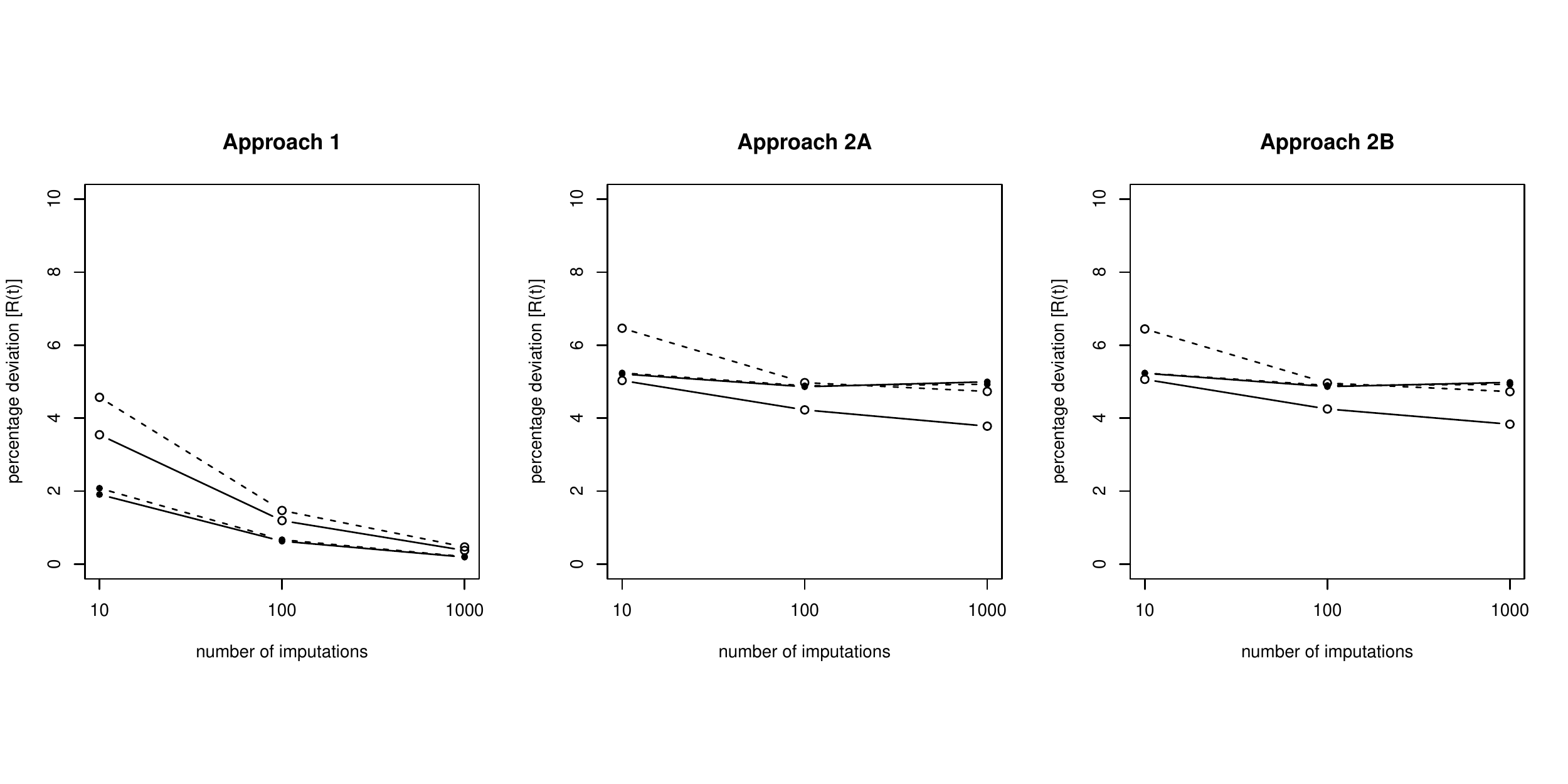}
\caption{\label{fig6}Percentage deviations of predictions $R(t)$ across replicate calibrations for approaches 1, 2 and 3 versus number of imputations at 1 and 5 years follow-up for the CRT data. Dashed lines are for $t=5$ and solid lines are for $t=1$. Solid filled dots corresponds to results for fully observed records and open dots are for predictions with partially missing records.}
\end{figure}

\begin{figure}[!htbh]
\centering
\includegraphics[height=7cm]{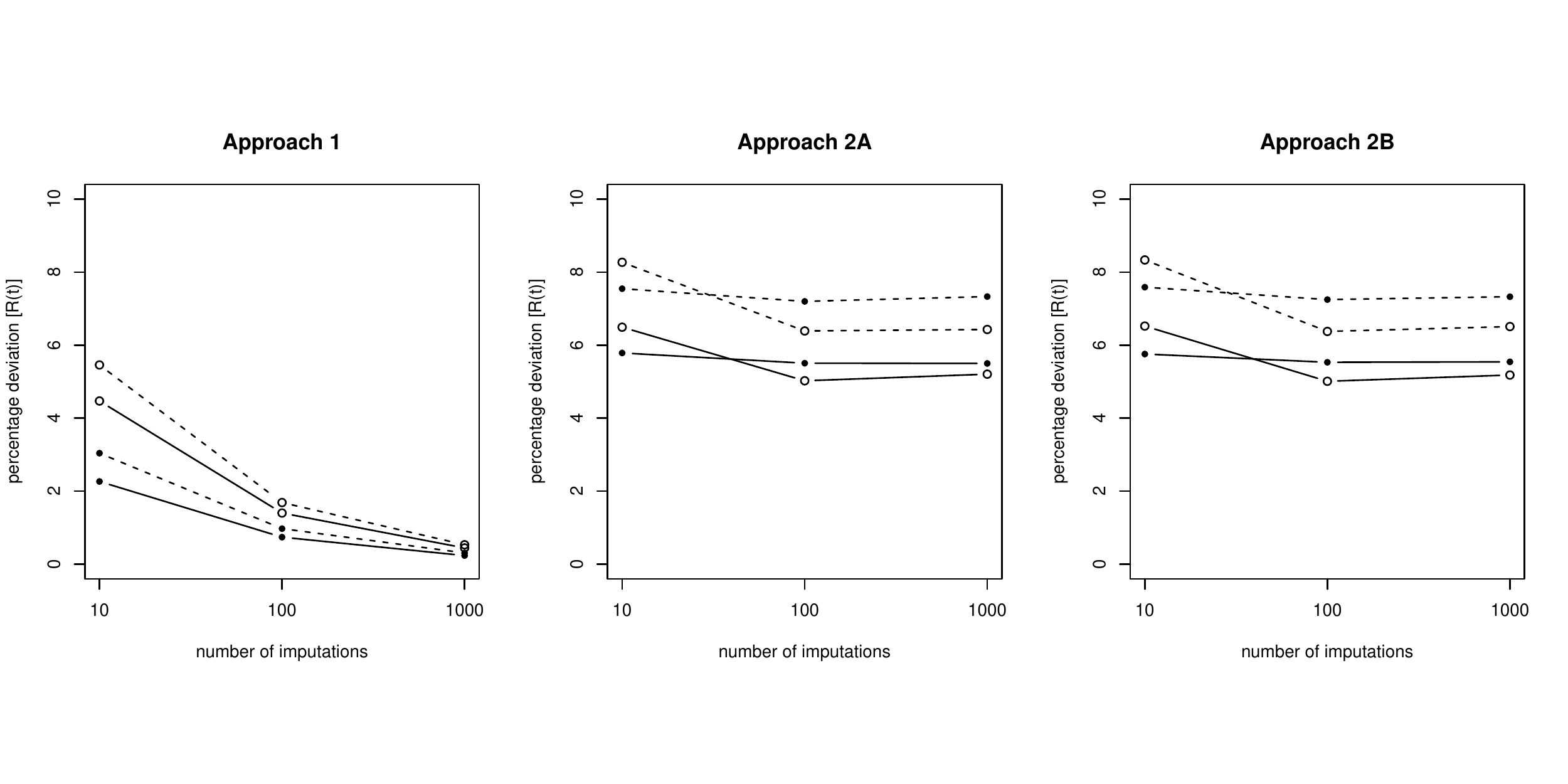}
\caption{\label{fig5}Percentage deviations of predictions $R(t)$ across replicate calibrations for approaches 1, 2 and 3 versus number of imputations at 1 and 5 years follow-up  for the CLL data. Dashed lines are for $t=5$ and solid lines are for $t=1$. Solid filled dots corresponds to results for fully observed records and open dots are for predictions with partially missing records.}
\end{figure}

Figures~\ref{fig5} and \ref{fig6} plot the calculated $R(t)$ measures (tables~\ref{table5} and \ref{table6}) versus number of imputations ($K=10,100,1000$) and for $t=1$ and $5$ years.  The difference between approach 1 versus 2A and 2B is evident.
Even at $K=10$, approach 1 reduces the variation measure `R' to levels which are not matched for approaches 2A and 2B with $K=1000$.
Reductions in predictive variation achieved by using multiple imputation as compared to single-imputation are substantial,  particulary for approach 1, which emphasizes the need for multiple imputation in prediction using large values of $K$.

Finally,  variation results calculated using the na\"ive implementations   point to slightly smaller variation measures.   From a qualitative  point of view,  conclusions match those from full implementations.

%

\subsection{Accuracy for the CRT and CLL data}
For the CRT and CLL datasets, we evaluate all methods (1, 2A and 2B, and the ``na\"ive" versions) using the Brier scores,  as defined in section 3.4 of \cite{Houwelingen2012}  at both 1 and 5 years follow-up, taking into account censoring.
The scores were calculated using the R package ``pec" \cite{Gerds2018} for each replicate (calibration) and then averages taken across replicates, for each value of  $K=1, 10, 100$ and 1000.

Table~\ref{table3}  shows results for $K=10$ at 1 and 5 years follow-up. We give summary measures on the full set of cross-validated predictions (column ``All obs.").  The calculation is then repeated using only the predictions corresponding to observations with missing values (``Missing") and likewise using only the completely observed observations (``Fully obs."). Note that the latter is not to be confused with so-called ``complete-case analysis", which would instead subset the patients based on the available information \emph{prior} to the analysis. In addition, this shows results based on 10-fold cross-validation and  single estimation of Cox models in the corresponding calibration sets for each left-out fold in a complete-case analysis,  without imputation.

%
%

	\begin{table}[ht]
	\centering
	\begin{tabular}{|r|rrr|rrr|}
		\hline
		\textbf{20 replicates} &

		\multicolumn{3}{c|}{\textbf{CRT data}} &
		\multicolumn{3}{c|}{\textbf{CLL data}}\\
		\textbf{K=10} &
		\textbf{Missing} & \textbf{All obs.} & \textbf{Fully obs.}&
		\textbf{Missing} & \textbf{All obs.} & \textbf{Fully obs.} \\
		\hline
		\multirow{5}{1.5cm}{\textbf{1 year}}
		\textbf{A 1 \ } & 0.071 & 0.067 & 0.063 & 0.202 & 0.189 & 0.182 \\
		\textbf{A 2A} & 0.071 & 0.067 & 0.063 & 0.202 & 0.189 & 0.182 \\
		\textbf{A 2B} & 0.071 & 0.067 & 0.063 & 0.202 & 0.189 & 0.182 \\
		\textbf{N 1 \ } & 0.070 & 0.066 & 0.063 & 0.196 & 0.186 & 0.181 \\
		\textbf{N 2A}& 0.070 & 0.067 & 0.063 & 0.198 & 0.187 & 0.181 \\
		\textbf{N 2B} & 0.070 & 0.067 & 0.063 & 0.198 & 0.187 & 0.181 \\
		\hline
		\textbf{Complete Case} & \ & \ & 0.063 & \ & \ & 0.184 \\
		\hline
		\multirow{5}{1.5cm}{\textbf{5 years}}
		\textbf{A 1 \ } & 0.211 & 0.191 & 0.176 & 0.237 & 0.241 & 0.242 \\
		\textbf{A 2A} & 0.211 & 0.191 & 0.177 & 0.237 & 0.242 & 0.243 \\
		\textbf{A 2B} & 0.211 & 0.191 & 0.177 & 0.237 & 0.242 & 0.243 \\
		\textbf{N 1 \ } & 0.204 & 0.187 & 0.176 & 0.229 & 0.236 & 0.239 \\
		\textbf{N 2A}& 0.205 & 0.188 & 0.176 & 0.230 & 0.237 & 0.240 \\
		\textbf{N 2B} & 0.205 & 0.188 & 0.176 & 0.230 & 0.237 & 0.240 \\
		\hline
		\textbf{Complete Case} & \ & \ & 0.178 & \ & \ & 0.256 \\
		\hline
	\end{tabular}
\caption{\normalsize\label{table3} Mean Brier score statistics across 20 replications for both the CRT and CLL data at 1 and 5 years follow-up,  based on combined  cross-validation and multiple imputation across approaches, using $K=10$ and $L=10$.}
\end{table}

\begin{figure}[!htbh]
\centering
\includegraphics[height=7cm]{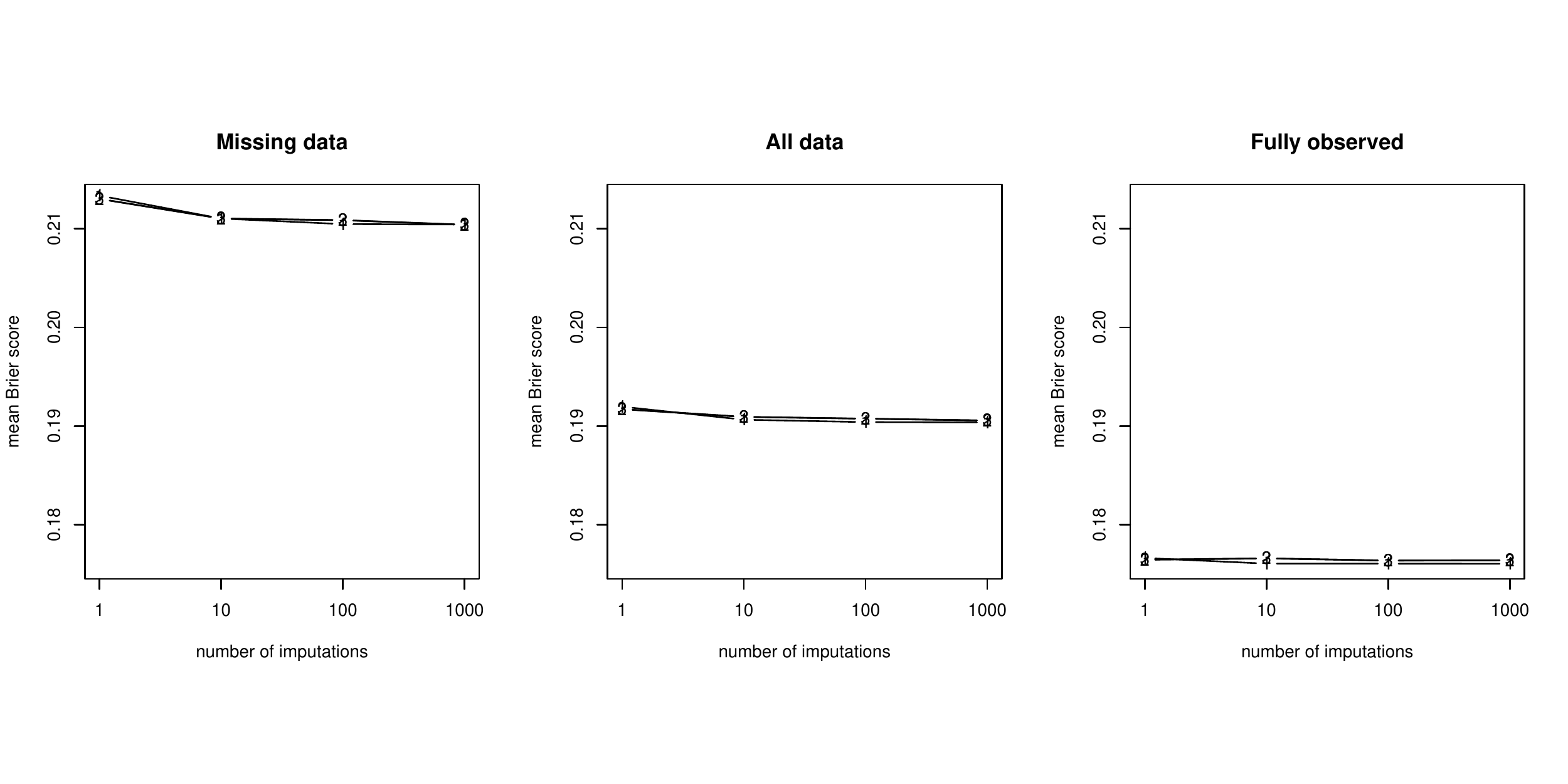}
\caption{\label{fig9} Mean Brier scores across 20 replicate analyses and at 5 years follow-up from approaches 1, 2A and 2B for the CRT data versus imputation number $K$. The plotting symbols (1, 2, 3) refer to methods 1, 2A and 2B in that order.}
\end{figure}

\begin{figure}[!htbh]
\centering
\includegraphics[height=7cm]{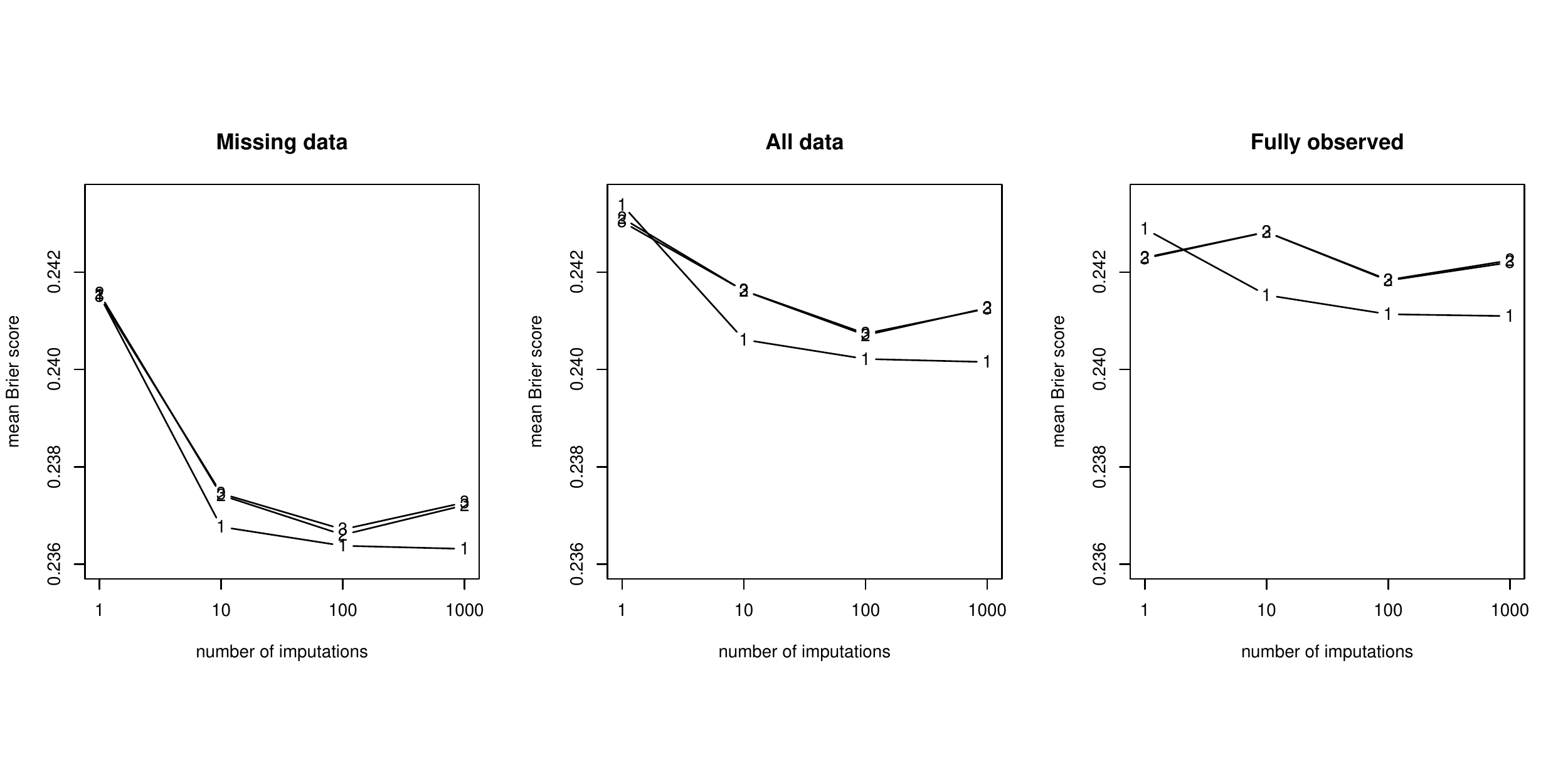}
\caption{\label{fig10}Mean Brier scores across 20 replicate analyses and at 5 years follow-up from approaches 1, 2A and 2B for the CLL data versus imputation number $K$. The plotting symbols (1, 2, 3) refer to methods 1, 2A and 2B in that order.}
\end{figure}

All methods have effectively equal performance, whether at 1 or 5 years follow-up. As expected, Brier scores are systematically larger at 5 years follow-up.
Results  calculated with the full set of predictions appear to be a mixture between the corresponding results calculated using either predictions on partially-missing or  completely observed cases only. Higher values may be expected on the missing data portion as a consequence of the increased uncertainty in prediction.   We note that the calculated Brier scores for the complete case analysis for the CRT data (0.0629 and 0.178) closely match results for the fully observed data across approaches.  This is not unsurprising,  as missing values are concentrated in a single predictor predominantly for this data and is due to failure of the measurement procedure,  which could reasonably be considered  an MCAR scenario. On the CLL data,  the complete case analysis yields a larger Brier score at 5 years,  which may be due to loss of efficiency due to reduced sample size.
The complementary tables for $K=100$ and 1000 are virtually indistinguishable from results for $K=10$ and hence not shown to save space.
The standard deviations calculated on the replicate Brier scores were systematically smaller than 0.005 and not shown to save space. Figures~\ref{fig9} and \ref{fig10}
show mean Brier scores versus imputation numbers $K$,  calculated separately on the full set of prediction (middle picture), on predictions for partially-missing records (left picture) and for fully observed records (right picture), for the CRT and CLL data separately. The pictures confirm the above discussion, but note that Brier scores are slightly larger for $K=1$ as compared to $K \ge 10$.
We finally note that results on the na\"ive methods tend to yield slightly smaller Brier scores,  which may be due to bias,  though the reduction is small.

\section{Simulations}
\subsection{Simulations for  imputation with survival outcome}
\subsubsection{Simulating lifetimes}
A disadvantage of the above data analytic applications is that we cannot assess bias as we do not know the true population survival probabilities.
We therefor study simulations. The below description of a simulation experiment  follows the setup previously described by \cite{Mertens2020} but adapted for  survival outcome with proportional hazards.
It is set up to simulate datasets resembling the CRT data to some extent.
For ease of reference, we repeat some of the simulation description from \cite{Mertens2020} and explain the modifications for survival analysis.

Simulated data is generated to consist of a survival time $T$, a censoring status indicator $\delta$ and a predictor matrix $\bm{X}$ consisting of $4$ continuous covariates which are drawn from a multivariate normal distribution $N_{4}(\bm{\mu},\bm{\Sigma})$, with $\mu_{j}=0$, $j=1,..,4$. For the covariance matrix $\bm{\Sigma}$ we choose the sample covariance matrix
of the standardized continuous variables (Age, Egfr, Hb, Lvef) in the CRT data (supplementary materials).
The survival times $T_{i}$, where $i$ denotes the $i^{th}$ individual, with $i=1,...,n$ and $n=1000$  are generated from an exponential density with  proportional hazards
	\[
h(t|\bm{x}_{i})=\lambda \exp\{ \bm{x}_{i} \bm{\beta}\}
\]
with $\bm{x}_i$ the corresponding row-vector of predictors	and fixed baseline hazard $\lambda=0.0073$.
The hazard ratios are chosen as
\[
\bm{\beta}^{'}=(\beta_{1},\log(1.2),\log(0.85),\log(0.75)),
\]
such that $\beta_{2},\beta_{3},\beta_{4}$ are fixed,  while we allow $\beta_{1}$ to vary across simulation scenarios as discussed below.
Censoring times are drawn from a uniform density on the interval $[13.5,167.5]$, resembling the CRT dataset.
The observed follow-up time $T$ is generated for each individual by taking the minimum between the simulated survival and censoring times, with status indicator $\delta$ set to 1 when $T$ corresponds to survival time and $0$ otherwise.  Finally, administrative censorship is applied to the simulated data at $t=84$ months, as for the CRT data. The above choices for $\lambda$ and the hazard ratios simulate  data with similar censoring and survival proportions as in the CRT data.

\subsubsection{Missing value scenarios}
Once the above described simulated outcomes and predictor data have been generated,  missing values are introduced by removing observations from predictor $X_1$, using 4  distinct scenarios. The scenarios are defined as all combinations of either low or high association of $X_1$ with survival outcome ($\beta_1=\log(1.1) \text{ or } \log(2)$)  and low or high numbers of missing observations (10 or 50 percent) in $X_1$ as indicated in table~\ref{table1}.
\begin{table}[htb]
\normalsize
\begin{center}
\begin{tabular}{|ll|c|c|}
\hline
&&\multicolumn{2}{|l|}{Percentage of missing observations ($X_1$)}\\
\cline{3-4}
&&Low [10\%] &High [50\%]\\
\hline
\multicolumn{1}{|l|}{log(hazard ratio) ($\beta_1$)}&Low [$\log(1.1$)]&Scenario 1& Scenario 3\\
\cline{2-4}
\multicolumn{1}{|l|}{                       }&High [$\log(2)$])&Scenario 2& Scenario 4\\ \hline
\end{tabular}
\caption{\normalsize\label{table1} Definition of 4 simulation scenarios in a $2^2$ design.  }
\end{center}
\end{table}
For each of these scenarios,  the missing values are introduced either {\it completely at random} (MCAR) or {\it at random} (MAR) (see \cite{Carpenter2013}, section 1.4 for definitions),  such that we have 8 scenarios in total. Again following \cite{Mertens2020}, for simulation subject to missing at random, unobserved values are introduced by calculating probabilities
$
p_{i}^{mar}=\min \bigl[{X_{2i}^{*}\ M}/{\overline{X}_{2}^{*}},1 \bigr],
$
for each $i^{th}$ individual, where $X_{2i}^{*}=(X_{2i}-\min(X_{2}))/(\max(X_{2})-\min(X_{2}))$  and $M$ denotes a fraction between 0 and 1 chosen to achieve the desired numbers of missing observations. A draw is then generated from the Bernoulli densities with probabilities $p_i^{mar}$ for each observation to decide on removal of the corresponding predictor values in $X_1$.

\subsubsection{Summary measures for simulations}
For each of the above described simulation scenarios,  we generate $100$ simulated datasets, with $n=1000$. For each simulated dataset, $20$ replications of each approach are generated. We  calculate average Brier scores $\overline{\overline{B}}$, averaging across both the simulations $s=1,...,100$ and replications $r=1,...,20$ for each scenario. In simulations, we may investigate bias as well as variance,  as we have the true survival fractions $S(t)_{iTrue}$ available for each $i^{th}$ simulated individual at any time $t$, based on the assumed simulation model. We assess these with the following definitions.

Let $\widehat{S}(t)_{irs}$ be the fitted survival probability from any approach at time $t$ and for the $i^{th}$ individual within the $s^{th}$ simulated dataset and for the $r^{th}$ replicate analysis.
We then calculate the mean $\overline{S(t)}_{is}$ across replications for the $i^{th}$ individual within the $s^{th}$ simulation and subsequently, the deviations $G(t)_{irs}=\widehat{S}(t)_{irs}-\overline{S(t)}_{is}$.
In analogy to the above description for the measures $D$, we now calculate the $90^{th}$ and $10^{th}$ percentiles $Q_{s_{0.9}}$ and $Q_{s_{0.10}}$ across all deviations $G(t)_{irs}$ corresponding for individuals within the 20$^{th}$ and 80$^{th}$ percentiles of $S(t)_{iTrue}$ and for the $s^{th}$ simulation.
 We next calculate the range statistics $R(t)_s=Q_{s_{0.9}}-Q_{s_{0.10}}$ as a measure of variation for each $s^{th}$ simulation and
 summarize using the average
$\overline{R(t)}$ across simulations $s$ as our final summary measure of variance.

To define a measure of bias,  we proceed similarly by first calculating the deviations of the average $\overline{S(t)}_{is}$ across replicates from the true survival fraction
\[
Bias(t)_{is}=\overline{S(t)}_{is}-S(t)_{iTrue}
\]
within the $s^{th}$ simulation.
We then calculate the mean $\overline{Bias(t)}_{s}$ deviation across i within $s$ and for those individuals with $0.2 \le S(t)_{iTrue} \le 0.8$.
We finally compute a summary measure of bias $\overline{\overline{Bias(t)}}$ by taking the average  of  the measures  $\overline{Bias(t)}_{s}$ across   $s$.

\subsection{Investigating simulations}
We investigate simulations as described in section 5.1, for both the MCAR and MAR scenarios and for $K=1, 10$ and $100$.  Results are presented for predictions at both 1 and 5 years follow-up and summarized in 12 tables (supplementary materials). These tables report the variance and bias measures discussed before as well as Brier score averages.
Tables give separate calculations of the statistics on predictions corresponding to either the fully observed records or for the records containing missing information.

Several aspects are of note.  First, variation in predicted probabilities is systematically larger for approaches 2A and 2B than for approach 1.  Likewise,  we find larger variations of predictions for observations with missing values than for fully observed observations.  The 5-year predictions are much more variable than the 1-year predictions.  Note that variability can be as large as 39\% ($K=10$, MCAR). This variability is reduced by increasing the number of imputations (maximum variability 5\% for $K=100$, MCAR). The `R'  variation measure can reduce substantially for the first approach when increasing numbers of imputations  to 100,  sometimes by up to an order of magnitude reduction. Variation improvements are much more resistant to increases in numbers of imputations for approaches 2A and 2B.  The results confirm and mirror those from the data analytic applications.

Bias tends to be small.   This holds both under MCAR and MAR. Approaches 1, 2A and 2B give very similar bias results under all scenarios. Bias is somewhat larger for the individuals with missing information than for those with full observations. It is typically below 2\%  and never above 4\% (absolute value). While some readers might find a 2\% bias non-negligible,  it should be noted that it is far smaller than the variance range levels we observed earlier in both data analysis and in these simulations also,  thus it appears a relatively benign problem. The bias can be both negative and positive. This might imply that including survival information through the Nelson-Aalen estimate  and censoring indicator may give reasonable calibrations for predictive purposes.

Brier scores seem smaller for patients with missing information than for those with full information
when na\"ive methods are used as compared to the full implementations. This may indicate these methods may give an optimistic view of the predictive potential of the data.

\section{Conclusions and Discussion}
This paper has investigated the generalization,  adaptation,  implementation and evaluation for  the analysis of censored survival outcome, of  methods presented in \cite{Mertens2020}  for binary outcome when imputations are used to account for missing predictor values.
Extension of these ideas for survival involves specific additional problems,  among which the need to adjust for censored outcomes in both model fitting as well as imputation.
For Cox regression specifically,  special care is needed to account for variation in estimated baseline hazards when combining imputation results.
In this paper we have restricted ourselves to Cox modeling throughout,  though the main ideas are completely general and  apply to any chosen survival modeling approach.

\subsection{Overview of key results}
Following \cite{Mertens2020},  we have described two basic approaches to survival prediction with Cox regression when imputations are used to account for missing values in either the calibration data or in the validation predictor measures. In both approaches,  the new `to be predicted' observations are added to the calibration set with outcomes deleted (as in cross-validation {\it e.g.}) or with the outcome data set to `missing'  (as in `true'  prediction) when not yet observed.   Multiple imputations are then generated on these augmented data matrices, imputing both missing calibration data as well as validation predictors (when missing).   This application of imputation for prediction  is in itself not new and has been described as concept before by at least \cite{Mercaldo2018} (section 1.2) for example,  though not implemented or evaluated. Models are then fit on the imputed calibration records and applied for prediction on the (possibly imputed) validation predictor data. As described in \cite{Mertens2020},  the two approaches consist of either averaging  the  predictions obtained from the distinct models fits on the imputed calibration sets (prediction averaging,  approach 1),  or generation of a pooled model first, which is subsequently applied to the (possibly imputed) validation predictor records (model pooling,  approach 2). For the second approach and Cox modeling specifically,  we have proposed two distinct implementations, defined as either a direct pooling of the distinct estimated baseline hazards or a re-estimation of the baseline based on pooled regression coefficients.

We investigated applications of proposed methods  in two novel real datasets.  In addition,  a simulation study was presented which generates lifetime outcome data subject to censoring and with missing data in predictors. We used these to assess the proposed methodologies.
The results from the validatory implementations for predictive Cox regression survival analysis reinforce conclusions drawn earlier based on the binary outcome analyses presented in \cite{Mertens2020}. We specifically confirmed the following substantive differences between methods in terms of the between-imputation variation (from replication of the analyses with different imputations),  while no meaningful differences or patterns were observed for accuracy.
\begin{enumerate}
  \item Single-imputation-based analysis should be avoided because of excessive levels of between-imputation variation. We note that the two approaches discussed (prediction averaging or use of a pooled model) coincide for single-imputation.
  \item Multiple imputations should be used in predictive calibration,  but prediction-averaging should be preferred as it gives greater reduction in between-imputation variation for the same numbers of imputations. From the two data-based analyses,  we can infer that prediction-averaging should be preferred {\it irrespective of the number of imputations used}, as even when increasing the number of imputations to 1000, the variability of Rubin's rule-based predictive calibration is outperformed by approach 1 when using 10 imputations only (figures~\ref{fig5} and \ref{fig6}) in these examples.
  \item Only with prediction-averaging could between-imputation variance be reduced towards zero,  with no further reduction in replication variation beyond 100 imputations when using pooled models.  Numbers of imputations in practical predictive survival (Cox) modelling may need to be drastically increased in practical application,  irrespective of the implementation chosen.
\end{enumerate}

In addition,  we found that for the second approach,  no substantive differences in predictive potential were found between either of the 2 implementations of pooling the baseline hazard. Between-imputation variation was substantially larger for prediction at 5 years follow-up as compared to predicting at 1 year.  We would expect this to reflect a general property of the methods in the survival context. As already shown for the binary outcome analyses referred to earlier,  we find for same reasons that prediction on fully observed records is more easy as compared to prediction based on records requiring imputations.

We also investigated na\"ive implementations,  which generate all imputations on the full datamatrix, including outcome data to-be-predicted on validation folds in the cross-validatory evaluations.   We tend to observe modest reductions in between-imputation variation as well as Brier scores (accuracy) as one might have expected. It is of course unclear how any such approach could be applied to `true prediction'   in situations where the outcome is as yet to be observed.
Na\"ive implementations may not be trustworthy and may exhibit optimistic bias.
These findings are in line with recent related literature  due to \cite{Wahl2016},  which may also be recommended to readers for an alternative viewpoint using bootstrap and verification with extensive simulation.

\subsection{Multiple imputation, Cox regression, prediction and cross-validation}
\subsubsection{Literature on Cox regression and imputation}
Published literature on imputation in the survival context is still somewhat limited.  Carpenter and Kenward (2013) dedicate a book section (8.1) to the topic,  while \cite{White2009} discuss imputation for the Cox model specifically. The Cox model is also considered as an example by \cite{Bartlett2015} in a paper on congenial imputation. Earlier contributions are due to \cite{vanBuuren1999} and \cite{Lin1993}.
 A more recent paper has investigated imputation in the context of time-dependent covariates with Cox regression \cite{Keogh2016}, while \cite{Bartlett2016} investigates imputation in competing risk settings. These publications investigate model and effect estimation. Wood, Royston and White (2005) \cite{Wood2015} study prediction problems with imputation and refers to ``extension to the Cox proportional hazards model"  as ``relatively straightforward" (page 616, section 2.1),  but without considering the survival application further nor providing  more detail.

\subsubsection{Cox regression prediction and imputation model choice}
To implement imputations with survival outcomes,  we have used the Nelson-Aalen cumulative hazard estimator and censoring indicator to replace the observed survival data as explained previously \cite{White2009}, \cite{Carpenter2013} (section 8.1.3, page 173). An advantage of the approach with survival data is that it can easily be implemented for survival data with standard imputation software,  such as the `\textit{MICE}' package we have purposely used for that reason in our analyses. Concerns have been raised about the effects of so-called `non-congeniality' of imputations implemented in this manner for survival analyses with the Cox model, specifically with respect to
potential bias. Bartlett et al. propose an alternative imputation strategy `SMC-FCS' in a Bayesian flavour to address these issues \cite{Bartlett2015}. We have not investigated this alternative approach for Cox regression,  first because of the additional computational overhead,  which is already high in the methods described here and the need to use special purpose software. Furthermore,  the primary concern with non-congeniality seems to us to be with bias of effect estimation,  while we consider prediction problems.  As we have shown bias or accuracy does not appear to be a major concern with the methods and implementations described in this paper. We therefor would not expect worthwhile improvement for our current objectives in the prediction context with these alternative imputation approaches,  but again stress that this aspect was not investigated.

Another imputation approach recently developed and already discussed by \cite{Mertens2020} could also be viewed as facilitating congeniality through the fully Bayesian paradigm by Erler et al. \cite{Erler2015}\cite{Erler2017} (implemented as an R package `\textit{jointAI}' \cite{Erler2020}).
While we recognize the obvious appeal, we are unsure to what extent Cox modeling could be accommodated with this software and especially for predictive purpose.
Cox modeling is complex in the Bayesian paradigm and requires special attention \cite{Ibrahim2001}.

Finally, it seems that while the notion of congeniality is of interest,  it is nevertheless concerned with `within-model'   coherence of imputations and specifically with respect to what in our application would be the final assumed prediction model.
Congeniality itself however does not ensure compatibility of the assumed model assumption with the underlying population distribution for which we wish to derive predictions.
We anticipate that the latter problem of how to study the credibility of modeling and imputation assumptions vis-a-vis the prediction target population is more important to the viability of practical prediction problems, with the congeniality problem a relatively minor issue. We also speculate that results presented in this paper would be largely unaffected,  even if congenial forms of imputation were used throughout.  Nevertheless, as these seem largely unexplored,  guidelines and experience for  practical data-based predictive analysis with imputation still need to be developed. This is outside the scope of present paper and must be left for future research.

%
%
%
%

\subsection{Towards practical clinical application}
We conclude with a discussion of the relation of this research to practical  application with an indication of research issues  still outstanding.

The first is that we have restricted ourselves to the scenario where the basic model is fully specified, including the set of predictors, without further uncertainty on the model definition. We note that the CRT data example is actually of this type,  as the original clinical application reported model fitting and assessment conditional on a fixed set of covariates. As discussed by \cite{Mertens2020},  but also \cite{Marshall},  a case could be made in favour of using pre-specified models in predictive model calibration,  to avoid the variation induced by model search procedures.
Of all the potential extensions to address model (predictor) uncertainty,  inclusion of shrinkage-based estimation would likely be the most interesting and promising in this context and combine more easily with the methods discussed in this paper. This is left for future research.

We have restricted to use of the Cox model,  though we expect conclusions to carry over broadly for other survival models too,  specifically accelerated failure time modeling, due to the highly non-linear nature of survival model prediction. For the Cox model,  treatment (and specifically pooling) of the baseline hazards function poses a special issue and we have investigated two pooling approaches.  These appear equivalent from a practical performance point of view.

As we have stressed a number of times,  the prediction-averaging approach appears to be highly competitive,  based on the assessment implemented in this paper.  This would suggest this method should be applied with imputations for prediction by preference. \cite{Mertens2020} already describe in their discussion that the pooled model may be of interest nevertheless, to give insight into the magnitudes and nature of the predictor effects driving the prediction. One could also imagine special prediction-type applications where predictions are required at some aggregate level,  such as a hospital or treatment unit for example,  which are themselves based on the individual patient-based predictions generated from models fit on existing patient records.  We cannot be sure that the preference for use of the averaging approach would be as strong in such problems,  although the relevance of a single pooled model could in such applications also be reduced. One could also imagine that reductions in the number of imputations required might be feasible in such scenarios.

Further advances in computing - but also statistical software - would also reduce the burden of maintaining several parallel prediction models.  Indeed parallel computing itself is highly suited to the prediction-averaging approach. \cite{Mertens2020} describes at more length the link between prediction-averaging and machine learning methodology.

We have not investigated the impact of (calibration) sample size on results.  While we would regard the current examples and simulations as of moderate sample sizes,  one could imagine clinical applications for much smaller experiments than those considered here.  In such circumstances one could imagine imputation variation to increase,  which would presumably increase differences between the prediction-averaging and pooled model approaches in prediction. Conversely,  perhaps the differences between the distinct methodological approaches investigated in this paper might reduce to the benefit of  using fully pooled models (approaches 2)  for truly large-sample problems,  but this would need to be verified in future research.

\mbox{}
\\
\mbox{}
\\
\noindent {\bf Acknowledgements}\\
\mbox{}
\\
 I thank Johannes Schetelig (University Hospital of the Technical University Dresden/DKMS Clinical Trials Unit, Dresden, Germany) for making available the CLL dataset. I thank Ulas Hoke and Nina Ajmone (Dep. Cardiovascular Research, LUMC, The Netherlands) for the opportunity and fruitful collaboration working on the CRT data. I also thank EBMT and DKMS for their work in collecting and preparing the CLL data. I acknowledge Liesbeth de Wreede for extensive discussion and John Zavrakidis, Ali Abakar Saleh and Theodor Balan for their valuable contributions.


\begin{thebibliography}{}

%

\harvarditem[Bartlett et. al.]{Bartlett}{2015}{Bartlett2015} Bartlett, J. W., Seaman, S. R., White, I. R. and Carpenter, J. R. (2015) Multiple Imputation of covariates by fully conditional specification: Accomodating the substantive model. {\it Statistical Methods in Medical Research}, {\bf 24(4)}, 462-487.

\harvarditem[Bartlett]{Bartlett and Taylor}{2016}{Bartlett2016} Bartlett, J. W. and Taylor, J. M. G. (2016) Missing covariates in competing risks analysis. {\it Biostatistics}, {\bf 17(4)}, 751-763.

\harvarditem[Carlin]{Carlin}{2015}{Carlin2015} Carlin, B. (2015) Multiple Imputation: a Perspective and Historical Overview. In: Handbook of Missing Data methodology. Molenberghs, G., Fitzmaurice, G., Kenward, M., Tsiatis, A., Verbeke, G. (Eds.) (2015) Chapman and Hall.
\harvarditem[Carpenter]{Carpenter}{2013}{Carpenter2013} Carpenter, J. and Kenward, M. (2013) Multiple Imputation and its Application. New York: Wiley.
\harvarditem[Clayton]{Clayton}{1994}{Clayton1994} Clayton, D. G. (1994) A Monte Carlo method for Bayesian inference in frailty models. Technical report, Medical Research Council Biostatistics Unit, Cambridge.


\bibitem[Erler, N. {\it et} {\it al.} (2016)]{Erler2015} Erler, N., Rizopoulos, D., van Rosmalen, J., Jaddoe, V., Franco, O., Lesaffre, E. (2016) Dealing with missing covariates in epidemiological studies: a comparison between multiple imputation and a full Bayesian approach.  \textit{Statistics in Medicine}, 35, 2955-2974.

\bibitem[Erler, N. {\it et} {\it al.} (2019)]{Erler2017} Erler, N., Rizopoulos, D., Jaddoe, V.,  J., Franco, O., Lesaffre, E. (2019) Bayesian imputation of time-varying covariates in linear mixed models.  \textit{Statistics Methods in Medical Research}, 28(2), 555-568.



\bibitem[Erler, N. {\it et} {\it al.} (2020)]{Erler2020} Erler, N. (2020) JointAI: Joint Analysis and Imputation of Incomplete Data {\it R package} \url{https://cran.r-project.org/web/packages/JointAI/JointAI.pdf}



\harvarditem[Gerds]{Gerds}{2018}{Gerds2018} Gerds, T.A. (2018) Prediction Error Curves for Risk Prediction Models in Survival
Analysis. {\it R package} \url{https://cran.r-project.org/web/packages/pec/pec.pdf}
\harvarditem[H\"oke et al.]{H\"oke, Mertens, Khidir, Schalij, Bax, Delgado and Marsan}{2017}{Hoke} H\"oke, U., Mertens, B., Khidir, M., Schalij, M., Bax, J., Delgado, V. and Marsan, N.  (2017) Usefulness of the CRT-SCORE for Shared Decision
Making in Cardiac Resynchronization Therapy in Patients With a Left Ventricular Ejection Fraction of $\le$ 35. {\it Am. J. Cardiology}, {\bf 120}, Issue 11, 2008-2016.

\harvarditem[Houwelingen and Putter]{Houwelingen and Putter}{2012}{Houwelingen2012} van Houwelingen , H.C. and Putter, H. (2012) {\it Dynamic Prediction in Clinical Survival Analysis.} Chapman and Hall/CRC Press.

\harvarditem[Ibrahim et al.]{Ibrahim, Chen and Sinha}{2001}{Ibrahim2001} Ibrahim, J.G., Chen, M.-H., Sinha, D. (2001) Bayesian Survival Analysis. Springer-Verlag.


\harvarditem[Keogh]{Keogh and Morris}{2016}{Keogh2016} Keogh, R. H. and Morris, T. P. (2016) Multiple imputation in Cox regression when there are time-varying effects of covariates. {\it Statistics in Medicine}, {\bf 37}, 3661-3678.




\harvarditem[Lin]{Lin}{1993}{Lin1993} Lin, D. Y. and Ying, Z. (1993) Cox regresion with incomplete covariate measurements.  {\it Journal of the American Statistical Association}, {\bf 84}, 42-52.

\harvarditem[Little]{Little}{1992}{Little1992} Little, R.J.A. (1992) Regression with missing Xs: A review. {\it Journal of the American Statistical Association}, {\bf 87}, 1227-1237.

\bibitem[Marshall, A.  {\it et} {\it al.} (2009)]{Marshall} Marshall, A., Altman, D., Holder, R. and Royston, P. (2009) Combining estimates of interest in prognostic modelling studies after multiple imputataion: current practice and guidelines. \textit{BMC Medical Research Methodology}, 9:57, doi:10.1186/1471-2288-9-57, \url{https://doi.org/10.1186/1471-2288-9-57}.

\harvarditem[Mercaldo and Blume]{Mercaldo and Blume}{2018}{Mercaldo2018} Mercaldo, S.F., Blume, J.D. (2018)  Missing data and prediction: the pattern submodel. {\it Biostatistics}, \url{https://doi.org/10.1093/biostatistics/kxy040}.

\harvarditem[Mertens et al.]{Mertens, Banzato and de Wreede}{2020}{Mertens2020} Mertens, B.J.A., Banzato, E. and de Wreede, L.C. (2020)  Construction and assessment of prediction rules for binary outcome in the presence of missing predictor data using multiple imputation: theoretical perspective and data-based evaluation. {\it Biometrical Journal}, , \url{https://doi.org/10.1002/bimj.201800289}.

\harvarditem[Moons]{Moons}{2006}{Moons2006} Moons, K.G.M., Donders, A.R.T., Stijnen, T. and Harrell, F.E. (2006) Using the outcome for imputation of missing predictor values was preferred. {\it Journal of Clinical Epidemiology}, {\bf 59}, 1092-1101.

\harvarditem[R Core Team]{R Core Team}{2019}{R2019} R Core Team (2019). R: A language and environment for statistical computing. R Foundation for Statistical Computing, Vienna, Austria. URL
  \url{https://www.R-project.org/}.

\harvarditem[Rubin]{Rubin}{1987}{Rubin1987} Rubin, D. B. (1987) Multiple Imputation for Nonresponse in Surveys. New York: Wiley.


\harvarditem[Schetelig et al.]{Schetelig, de Wreede and van Gelder}{2017a}{Schetelig2017a} Schetelig, J., de Wreede, L.C. , van Gelder, M. {\it et al.} (2017a) Risk factors for treatment failure after allogeneic transplantation of patients with CLL: a report from the European Society for Blood and Marrow Transplantation. {\it Bone Marrow Transplantation},  52, 552-560

\harvarditem[Schetelig et al.]{Schetelig, de Wreede and Andersen}{2017b}{Schetelig2017b} Schetelig, J., de Wreede, L.C. ,  Andersen, N.S., {\it et al.} (2017b) Centre characteristics and procedure-related factors have an impact on outcomes of allogeneic transplantation for patients
with CLL: a retrospective analysis from the European Society for Blood and Marrow Transplantation (EBMT). {\it British Journal of Haematology},  178, 521-533





\harvarditem[Spiegelhalter et al.]{Spiegelhalter, Thomas and Best}{1996}{Spiegelhalter1996} Spiegelhalter, D., Thomas, A., Best, N. (1996) BUGS 0.5. Bayesian inference Using Gibbs Sampling Manual (version ii). Technical report, Medical Research council Biostatistics Unit, Cambridge.


\harvarditem[Therneau and Lumley]{Therneau and Lumley}{2019}{Therneau2019} Therneau, T.M. and Lumley, T. (2019). Package `survival'.
  \url{https://CRAN.R-project.org/package=survival}.

\harvarditem[vanBuuren]{vanBuuren}{1999}{vanBuuren1999} van Buuren, S., Boshuizen, H. C. and Knook, D. L.  (1999) Multiple imputation of missing blood pressure covariates in survival analysis. {\it Statistics in Medicine}, {\bf 18}, 681-694.


\harvarditem[vanBuuren]{vanBuuren}{2015}{vanBuuren2015} van Buuren, S. (2015) Fully Conditional Specification. In: Handbook of Missing Data methodology. Molenberghs, G., Fitzmaurice, G., Kenward, M., Tsiatis, A., Verbeke, G. (Eds.) (2015) Chapman and Hall.


\harvarditem[van Buuren and Groothuis-Oudshoorn]{van Buuren and Groothuis-Oudshoorn}{2011}{vanBuuren2011} van Buuren S, Groothuis-Oudshoorn K (2011). “mice: Multivariate Imputation by Chained Equations in R.” Journal of Statistical Software, 45(3), 1-67, \url{https://www.jstatsoft.org/v45/i03/}.





\harvarditem[Wahl et al.]{Wahl, Boulesteix, Zierer, Thorand and van de Wiel}{2016}{Wahl2016} Wahl, I., Boulesteix, A.-L., Zierer, A., Thorand, B. and van de Wiel, M. (2016) Assessment of predictive performance in incomplete data by combining internal validation and multiple imputation. {\it BMC Medical Research Methodology}, 16(1):144.

\harvarditem[White and Royston]{White and Royston}{2009}{White2009} White, I. R. and Royston, P.  (2009) Imputing missing covariate values for the Cox model. {\it Statistics in Medicine}, {\bf 28}, 1982-1998.

\bibitem[Wood, A. {\it et} {\it al.} (2015)]{Wood2015} Wood, A., Royston, P. and White, I. (2015) The estimation and use of predictions for the assessment of model performance using large samples with multiply imputed data. \textit{Biometrical Journal}, 57, 614-632.


\end{thebibliography}
\end{document}